\documentclass[journal,twocolumn,10pt]{IEEEtran}
\IEEEoverridecommandlockouts
\usepackage[T1]{fontenc}
\usepackage{cite}
\usepackage[cmex10]{amsmath}
\usepackage{amsthm}
\usepackage{amssymb}
\usepackage{bbm}
\usepackage{mathtools}
\usepackage[hidelinks]{hyperref}
\usepackage{algorithmic}
\usepackage{textcomp}
\usepackage[usenames, dvipsnames]{color}
\usepackage[linesnumbered,ruled]{algorithm2e}
\usepackage{subcaption}
\usepackage{graphicx}  
\usepackage{url}
\usepackage[usestackEOL]{stackengine}
\usepackage{verbatim}
\usepackage{dsfont}

\usepackage {tikz}
\tikzset{>=stealth}
\usepackage{xcolor,colortbl}
\definecolor{lightred}{RGB}{255, 240, 240}
\definecolor{lightblue}{RGB}{230, 250, 255}
\definecolor{lightgreen}{RGB}{240, 255, 242}
\definecolor{myred}{RGB}{220, 0, 0}
\definecolor{myblue}{RGB}{0, 17, 173}
\definecolor{mygreen}{RGB}{2, 117, 0}

\newcommand{\mca}[1]{\mathcal{#1}}

\newcommand{\mbf}[1]{\mathbf{#1}}
\newcommand{\mbb}[1]{\mathbb{#1}}
\newcommand{\mrm}[1]{\mathrm{#1}}
\newcommand{\mtt}[1]{\mathtt{#1}}

\newcommand{\Lb}{L_{\mathrm{b}}}
\newcommand{\Lc}{L_{\mathrm{c}}}
\newcommand{\Ls}{L_{\mathrm{s}}}
\newcommand{\Lsb}{L_{\mathrm{sb}}}
\newcommand{\Nsb}{N_{\mathrm{sb}}}
\newcommand{\ec}{\epsilon_{\mathrm{c}}}

\newcommand{\op}[1]{\operatorname{#1}}
\newcommand{\mymod}[1]{\mathrm{mod}}
\DeclareMathOperator*{\argmin}{argmin}

\DeclareMathOperator*{\maximize}{maximize}
\DeclareMathOperator*{\minimize}{minimize}
\DeclareMathOperator*{\st}{subject\;to}

\makeatletter
\renewcommand*\env@matrix[1][\arraystretch]{%
  \edef\arraystretch{#1}%
  \hskip -\arraycolsep
  \let\@ifnextchar\new@ifnextchar
  \array{*\c@MaxMatrixCols c}}
\makeatother

\ifCLASSINFOpdf
\else
\fi
\usepackage{amsmath}
\begin{document}
%
\title{Channel-Coded Precoding for Multi-User MISO~Systems}

\author{
Ly~V.~Nguyen, Junil Choi, Bj\"{o}rn Ottersten, and A.~Lee~Swindlehurst
\thanks{This work was supported by U.S. National Science Foundation Grant CCF-2008724.}
\thanks{Ly V. Nguyen and A. Lee Swindlehurst are with the Center for Pervasive Communications and Computing, Henry Samueli School of Engineering, University of California, Irvine, CA, USA 92697 (e-mail: vanln1@uci.edu, swindle@uci.edu).}
\thanks{Junil Choi is with the School of Electrical Engineering, KAIST, Daejeon, 34141, South Korea (e-mail: junil@kaist.ac.kr).}
\thanks{Bj\"{o}rn Ottersten is with the Interdisciplinary Center for Security, Reliability and Trust (SnT), University of Luxembourg, 1855 Luxembourg City, Luxembourg (e-mail: bjorn.ottersten@uni.lu)}
}

\maketitle

\begin{abstract}
Precoding is a critical and long-standing technique in multi-user communication systems. However, the majority of existing precoding methods {do not} consider channel coding in their designs. In this paper, we consider the precoding problem in multi-user multiple-input single-output (MISO) systems, incorporating channel coding into the design. {By leveraging the error-correcting capability of channel codes we increase the degrees of freedom in the transmit signal design, thereby enhancing the overall system performance.} We first propose a novel {data-dependent} precoding framework for coded MISO  systems, referred to as \textit{channel-coded precoding} (CCP), {which maximizes} the probability that information bits can be correctly recovered by the channel decoder. This proposed CCP framework allows the transmit signals to produce data symbol errors {at the users' receivers}, as long as the overall information BER performance can be improved. We develop the CCP framework for both one-bit and multi-bit error-correcting capacity and devise a projected gradient-based approach to solve the design problem. We also develop a robust CCP framework for the case where knowledge of perfect channel state information (CSI) is unavailable {at the transmitter}, taking into account the effect of both noise and channel estimation errors. Finally, we conduct numerous simulations to verify the effectiveness of the proposed CCP and its superiority compared to existing precoding methods, and we {identify situations} where the proposed CCP yields the most significant gains.
\end{abstract}


%
\IEEEpeerreviewmaketitle

\section{Introduction}
\label{sec:introduction}
Interference is unavoidable in wireless communications due to the broadcast nature of electromagnetic waves. Therefore, interference management (IM) is critical in ensuring a desired Quality of Service (QoS) in wireless networks. {Precoding is an effective IM technique that allows a base station (BS) or access point (AP) equipped with multiple antennas and accurate channel state information (CSI) to simultaneously manipulate the interference for multiple data streams and multiple users, each often equipped with a single antenna, thereby improving signal quality and system performance.} Precoding has been studied intensively in the literature. Dirty paper coding (DPC) is known to be theoretically capable of achieving the channel capacity~\cite{Costa1983DPC}. However, implementing DPC is challenging due to its high computational complexity and {sensitivity to CSI errors}. A practical and efficient alternative is linear precoding as it has low complexity and still can deliver good QoS.

Linear precoders view interference as a negative factor that must be mitigated as much as possible. Thus, they are typically designed to minimize the effect of noise and interference. Well-known examples of linear precoding include maximal ratio transmission (MRT), zero forcing (ZF), and minimum mean-squared error (MMSE) precoders~\cite{verdu1998multiuser}. The MRT precoder maximizes the signal-to-noise ratio (SNR) of the users~\cite{Price1958Rake} but makes no attempt to mitigate multi-user interference (MUI), and thus results in poor signal-to-interference-plus-noise ratio (SINR). Ideally, the ZF precoder completely cancels MUI essentially via channel inversion~\cite{Shnidman1967ZF,Caire2003ZF}, and thus significantly outperforms the MRT precoder. The MMSE\footnote{The MMSE precoder is referred to by other names in the literature such as regularized ZF~\cite{Peel2005Vector} and transmit Wiener filter~\cite{Joham2005Linear}.} precoder~\cite{Vojcic1998Transmitter,Joham2005Linear,Peel2005Vector} also relies on channel inversion, but it also considers the effect of the noise at the receivers in order to achieve an efficient balance between interference mitigation and noise reduction. This enables the MMSE precoder to perform better than both the ZF and MRT precoders. 

Interestingly, the MMSE precoder turns out to be a special case of the optimal solution to the problem of designing a precoder to maximize the 
SINR~\cite{Emil2014Optimal}, an approach that was first suggested as a generalized Rayleigh quotient in~\cite{Zetterberg1995Spectrum}. The power-constrained SINR maximization problem was later formulated as a standard conic program in~\cite{Wiesel2006Linear}. A similar problem was also studied in~\cite{Dartmann2013Duality} where both per-antenna and per-antenna-array power constraints were considered. Another problem of significant interest is the SINR-constrained power minimization problem, which was considered in~\cite{Rashid1998Transmit,Visotsky1999Optimum,Rashid1998Joint,Schubert2004Solution,bengtsson2018optimum,Wiesel2006Linear,Zhi2006Introduction}. The work in~\cite{Rashid1998Transmit,Visotsky1999Optimum,Rashid1998Joint,Schubert2004Solution} exploited the property of uplink-downlink duality to develop efficient iterative precoding algorithms, while~\cite{bengtsson2018optimum,Wiesel2006Linear,Zhi2006Introduction} showed that this problem can be relaxed or transformed into a convex optimization problem through semidefinite programming (SDP) and second-order cone programming (SOCP), {providing the optimal linear precoder in terms of SINR at the receivers.}

Recently, symbol-level precoding (SLP) has emerged as an advanced technique that can offer substantial performance improvements over traditional linear precoding methods. SLP is motivated by the observation that not all interference is harmful, and it exploits constructive interference (CI) to increase the effective power of the received signals by pushing them farther away from the symbol decision boundaries~\cite{Masouros2015Exploiting}. Early studies on SLP were conducted in~\cite{Masouros2007Novel,Masouros2009Dynamic,Masouros2011Correlation,Masouros2012THP,Garcia2014Power}. The approaches presentedin~\cite{Masouros2007Novel,Masouros2009Dynamic,Masouros2011Correlation} decompose the interference into two components including CI and destructive interference (DI). In~\cite{Masouros2007Novel} and~\cite{Masouros2009Dynamic}, the CI component is retained while the DI is canceled. In contrast, the method in~\cite{Masouros2011Correlation} does not cancel the DI component but transforms it into CI through a correlation-rotation approach. Other early CI-based precoding approaches were developed in~\cite{Masouros2012THP} and~\cite{Garcia2014Power} based on the Tomlinson-Harashima precoding technique. Unlike the approach in~\cite{Masouros2012THP} which optimizes only one scaling parameter for the first user, \cite{Garcia2014Power} {further improves performance} by simultaneously optimizing multiple scaling parameters for multiple users.

Since the work in~\cite{Masouros2015Exploiting}, SLP has attracted significant attention, leading to the development of more advanced and efficient SLP approaches~\cite{Alodeh2018Symbol,Li2020Tutorial}. For example, different CI-based concepts such as strict phase rotation, relaxed detection regions, and constructive symbol regions were developed for phase-shift keying (PSK) constellations in~\cite{Alodeh2015Constructive},~\cite{Alodeh2016Energy,Masouros2015Exploiting}, respectively, where the term ``constructive symbol region'' was later referred to as ``non-strict phase rotation'' in~\cite{Li2018Practical}. While strict phase rotation requires the noiseless received signals to have the same phase as the desired data symbols, the relaxed detection region approach allows for phase differences that remain within a certain phase margin. The constructive symbol region approach also permits deviations of the received signal from the desired data symbols, but the noiseless received signals must fall within a region determined by a given safety margin distance instead of a phase margin. For quadrature amplitude modulation (QAM), SLP designs based on symbol scaling were proposed in \cite{Alodeh2017MultiLevel,Li2021MultiLevel,Junwen2023Speeding} and shown to lead to a succinct and easy-to-handle optimization formulation. SLP for generic constellations was studied in~\cite{Haqiqatnejad2018Generic,Haqiqatnejad2018Distance} using an optimization metric based on the concept of a 
distance-preserving CI region. To reduce the computational complexity of SLP designs, numerous efforts have been made using closed-form sub-optimal solutions~\cite{Haqiqatnejad2018Power,Haqiqatnejad2019Approximate}, structured iterative algorithms~\cite{Li2018Practical,Li2021MultiLevel}, block-level optimization~\cite{Li2023Practical,Junwen2024Block}, user grouping~\cite{Zichao2022Low}, parallel processing~\cite{Junwen2024Low,Junwen2023Speeding}, and deep unfolding~\cite{Junwen2024ADMM}. CI-basesd SLP has also been applied in other scenarios such as for constant-envelope digital signals~\cite{Jedda2018Quantized,Tsinos2020ConstantEn}, 1-bit digital-to-analog converters (DAC)~\cite{Li2020Interference,Li2021Onebit,Zheyu2024Efficient,Ly2023SSP}, reconfigurable intelligent surfaces (RIS)~\cite{Li2020RIS,Rang2021SLP,Rang2021RIS}, and integrated sensing and communications (ISAC)~\cite{Rang2021DRFCSLP,Ly2024DFRCSLP,Yunwang2024ISACSLP}.

A limitation common to of all the aforementioned precoding designs is that they do not exploit redundancy in the transmitted signals due to channel coding. Conventional precoder designs typically focus performance metrics that will reduce or potentially minimize the symbol error rate (SER) of the already channel-encoded bits, rather than focusing on the underlying information bits that are the inputs to the channel encoder. While reducing the SER of the coded data symbols will in turn reduce the BER of the information bits, this design strategy {leaves untouched the potential to improve performance by incorporating knowledge of channel coder in the precoding design}. It is ultimately the BER of the information bits that is most important to overall system performance, not the SER of the coded data symbols. There is limited prior work in the literature that takes channel coding into account in the precoding design. For example, \cite{Jevgenij2023End2End} examined the performance of SLP in a coded system using the conventional log-likelihood ratio (LLR) and low-density parity check (LDPC) codes, demonstrating that SLP can still significantly outperform ZF. In~\cite{Yafei2024Soft}, the authors proposed to use a Gaussian mixture model to estimate the LLRs of non-Gaussian signals received after SLP. However, to the best of our knowledge, no prior work has addressed transmit signal design\footnote{In this paper, we refer the terms ``precoding'' and ``transmit signal design'' to as the same concept and use them interchangeably.} taking into account the effect of channel coding.

In this paper we consider the precoding problem in multi-user coded MISO systems and address the question ``{How can knowledge of channel coding be exploited for precoding designs in coded systems?}'' Our idea is to leverage the error-correcting capability of channel codes to increase the number of available degrees of freedom in transmit signal designs, thereby enhancing the overall system performance. The contributions of the paper are summarized as follows:
\begin{itemize}
    \item We first propose a novel precoding framework for coded MISO systems, referred to as \textit{channel-coded precoding} (CCP). The objective of the proposed CCP framework is to minimize the information BER. However, directly minimizing the information BER is challenging due to the {complexity} of the channel coding process. To overcome this, we take the novel approach of allowing the transmit signals to produce correctable data symbol errors at the receiver, as long as the overall information BER performance can be improved. {This precoding formulation introduces a novel design methodology} in which the channel coding and transmit signal designs could be jointly optimized.
    \item We study the CCP design problem under a transmit power constraint and consider the case where the error-correcting capacity of the channel code is one bit. A closed-form expression for the probability of correctly recovering the information bits for various QAM constellations, including 4-QAM, 8-QAM, and 16-QAM, is derived. We also develop an efficient projected gradient (PG)-based algorithm {to optimize our CCP design criterion.}
    \item We then extend our work to the case where the channel code has a multi-bit error-correcting capacity. Due to the complicated form of the probability of successfully recovering the information bits, we propose to divide the entire transmission block into multiple sub-blocks where we maximize the probability that there is no more than one bit error in each sub-block. This approach allows us to simplify the optimization problem and {more efficiently} design the transmit signals.
    \item We further study the case where only imprecise CSI is available and dwe evelop a robust CCP framework, taking into account the effect of both noise and channel estimation errors. We show that the proposed robust CCP approach outperforms other precoding methods.
    \item Finally, we conduct several simulation studies to verify the effectiveness of the proposed CCP framework and its superiority compared to existing conventional precoding methods. Based on the numerical results, we also suggest system settings where the proposed CCP yields the most significant gains.
\end{itemize}

The remainder of the paper is organized as follows. Section~\ref{sec:system_model} describes the system model, and in Section~\ref{sec:proposed_CCP_frame}, we briefly revisit conventional precoding methods before introducing our proposed CCP framework. The proposed CCP designs for the case of one-bit and multi-bit error-correcting capacity are developed in Section~\ref{sec:CCP_ec1} and Section~\ref{sec:CCP_ecg1}, respectively. Section~\ref{sec:robust_CCP} presents the proposed robust CCP design. Finally, we provide simulation results in Section~\ref{sec:numerical_results} and conclude the paper in Section~\ref{sec:conclusion}.

\textit{Notation:} Upper-case and lower-case boldface letters respectively denote matrices and column vectors. The $\ell_0$-norm and $\ell_2$-norm of a vector and the Frobenius-norm of a matrix are represented by $\|\cdot\|_0$, $\|\cdot\|_2$, $\|\cdot\|_F$, respectively. The transpose and conjugate transpose are respectively denoted by $[\cdot]^T$ and $[\cdot]^H$. The notation $\Re\{\cdot\}$ and $\Im\{\cdot\}$ denotes the real and imaginary parts of the complex argument, respectively. If $\Re\{\cdot\}$ and $\Im\{\cdot\}$ are applied to a matrix or vector, they are applied individually to every element of that matrix or vector. The unit imaginary number is denoted by $\mtt{j}$, satisfying $\mtt{j}^2=-1$. The functions $\mathcal{N}(\cdot,\cdot)$ and $\mathcal{CN}(\cdot,\cdot)$ respectively represent the normal distribution for real and complex random variables, where the first argument is the mean and the second argument is the variance or the covariance matrix. The cumulative distribution function of the standard Gaussian random variable is denoted by $\Phi(\cdot)$ and given as $\Phi(t) = \int_{-\infty}^{t}\frac{1}{\sqrt{2\pi}}e^{-\frac{\tau^2}{2}}d\tau$.

\section{System Model}
\label{sec:system_model}
We consider a coded downlink MISO system where a BS equipped with $N$ antennas simultaneously serves $K$ single-antenna users using the same time and frequency resource. The BS performs precoding to transmit a different data stream to each of the $K$ users. For each user $k$, a sequence of information bits $\mbf{b}_k = [b_{k,1},\, \ldots,\, b_{k,\Lb}]^T$ is encoded via a channel encoder, denoted as $\mca{C}(\cdot)$, to produce a sequence of coded bits $\mbf{c}_k = [c_{k,1},\, \ldots,\, c_{k,\Lc}]^T$, i.e., $\mbf{c}_k = \mca{C}(\mbf{b}_k)$, with code rate $r = \Lb/\Lc$. For simplicity, we assume the same channel encoder for all $K$ users. The sequence of data symbols for user $k$ is then $s_{k,1},\, \ldots,\, s_{k,\Ls}$, where $\Ls = \Lc/\log_2(M)$ is the number of data symbols and $M$ is the constellation size {which is assumed to be the same for all users.}

We assume the channel $\mbf{H} = [\mbf{h}^T_1,\,\ldots,\,\mbf{h}^T_K] \in \mbb{C}^{K \times N}$ is block flat-fading. Initially we assume that $\mbf{H}$ is known perfectly at the BS, but this assumption will be relaxed later. Let $\mbf{x}_t$ denote the transmit signal vector at time slot $t \in \{1,\,\ldots,\,\Ls\}$, and {let $\mbf{y}_t \in \mbb{C}^{K\times 1}$ be the corresponding received signal vector whose $k$-th element is the signal received by user $k$. The vector $\mbf{y}_t$ is given by}
\begin{equation}
    \mbf{y}_t = \mbf{H}\mbf{x}_{t} + \mbf{n}_t,
    \label{eq_sys_model}
\end{equation}
where $\mbb{E}[\|\mbf{x}_t\|^2] = \rho$ is the transmit power and $\mbf{n}_t\sim\mca{CN}(0,\sigma^2\mbf{I}_{K})$ is noise. {The elements of $\mbf{n}_t$ are assumed to have equal variance $\sigma^2$, which is known by the BS.} Let $\mbf{s}_t = [s_{1,t},\,\ldots,\,s_{K,t}]^T$ be the vector of data symbols at a time slot $t$. For convenience in later derivations, we denote the matrix of data symbols and transmit signals by $\mbf{S} = [\mbf{s}_1,\,\ldots,\,\mbf{s}_{\Ls}]$ and $\mbf{X} = [\mbf{x}_1,\,\ldots,\,\mbf{x}_{\Ls}]$, respectively.

\begin{figure}[t!]
    \centering
    \includegraphics[width=0.95\linewidth]{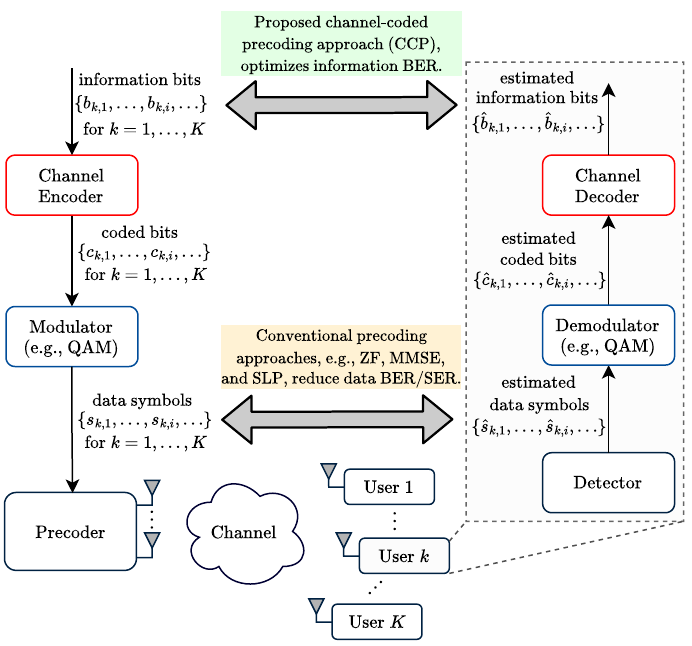}
    \caption{Proposed CCP versus conventional precoding approaches.}
    \label{fig:CCP}
\end{figure}

At the user, a detector is used to first estimate the transmitted data symbols $\mbf{\hat{S}} = [\mbf{\hat{s}}_1,\,\ldots,\,\mbf{\hat{s}}_{\Ls}]$ which are then passed through a demodulator to recover the transmitted coded bits $\mbf{\hat{c}}_k = [\hat{c}_{k,1},\, \ldots,\, \hat{c}_{k,\Lc}]^T$. Finally, the estimated information bits $\mbf{\hat{b}}_k = [\hat{b}_{k,1},\, \ldots,\, \hat{b}_{k,\Lb}]^T$ are recovered by a channel decoder as $\mbf{\hat{b}}_k = \mca{C}^{-1}(\mbf{\hat{c}}_k)$, where $\mca{C}^{-1}(\cdot)$ denotes the inverse of the channel encoder $\mca{C}(\cdot)$ used at the BS. The goal of the precoder is to design the transmit signal $\mbf{X}$ in order to facilitate the detection and decoding process at the receiver. Fig.~\ref{fig:CCP} gives a block diagram that illustrates the transmit and receive processes.

\section{Conventional Approaches vs. Proposed CCP}
\label{sec:proposed_CCP_frame}
In this section, we first briefly revisit the conventional precoding framework that has been studied {extensively} and highlight several prominent approaches in the literature. We then present our proposed CCP framework and show how the error-correcting capability of channel codes can be exploited for efficient precoding designs.

\subsection{Conventional Precoding Framework}
\label{subsec:conventional_precoders}
Conventional precoding can be viewed as mapping a data symbol matrix $\mbf{S}$ to a transmit signal matrix $\mbf{X}$ based on the CSI $\mbf{H}$:
\begin{equation}
    \mbf{X} = \mca{P}(\mbf{H},\mbf{S})
    \label{eq:conventional_precoding_frame}
\end{equation}
where $\mca{P}(\cdot)$ is the to-be-designed operator. The objective of precoding is to design $\mca{P}(\cdot)$ such that the recovered data symbol matrix $\mbf{\hat{S}}$ is ``close'' to $\mbf{S}$ in some sense and can be recovered as accurately as possible. Given the goal of optimizing the estimation of the data symbols, conventional approaches can be viewed as \textit{data-directed} precoders. As introduced earlier, linear precoding is a popular conventional precoding technique for which the mapping $\mca{P}$ is linear, i.e., $\mbf{X} = \mca{P}(\mbf{H},\mbf{S}) = \mbf{P}\mbf{S}$ for some matrix $\mbf{P}$ that is designed solely based on the channel matrix $\mbf{H}$. Typical linear precoders include the MRT, ZF, and MMSE approaches, for which $\mbf{P}$ is respectively given by
\begin{align*}
    &\mbf{P}_{\mtt{MRT}} = \beta_{\mtt{MRT}} \mbf{H}^H,\\
    &\mbf{P}_{\mtt{ZF}} = \beta_{\mtt{ZF}} \mbf{H}^H(\mbf{H}\mbf{H}^H)^{-1},\\
    &\mbf{P}_{\mtt{MMSE}} = \beta_{\mtt{MMSE}} (\mbf{H}^H\mbf{H} + K\sigma^2/\rho\mbf{I})^{-1}\mbf{H}^H,
\end{align*}
where $\beta_{\mtt{MRT}}$, $\beta_{\mtt{ZF}}$, and $\beta_{\mtt{MMSE}}$ are the scaling factors that control the transmit power.

\begin{figure}[t!]
    \centering
    \includegraphics[width=0.48\linewidth]{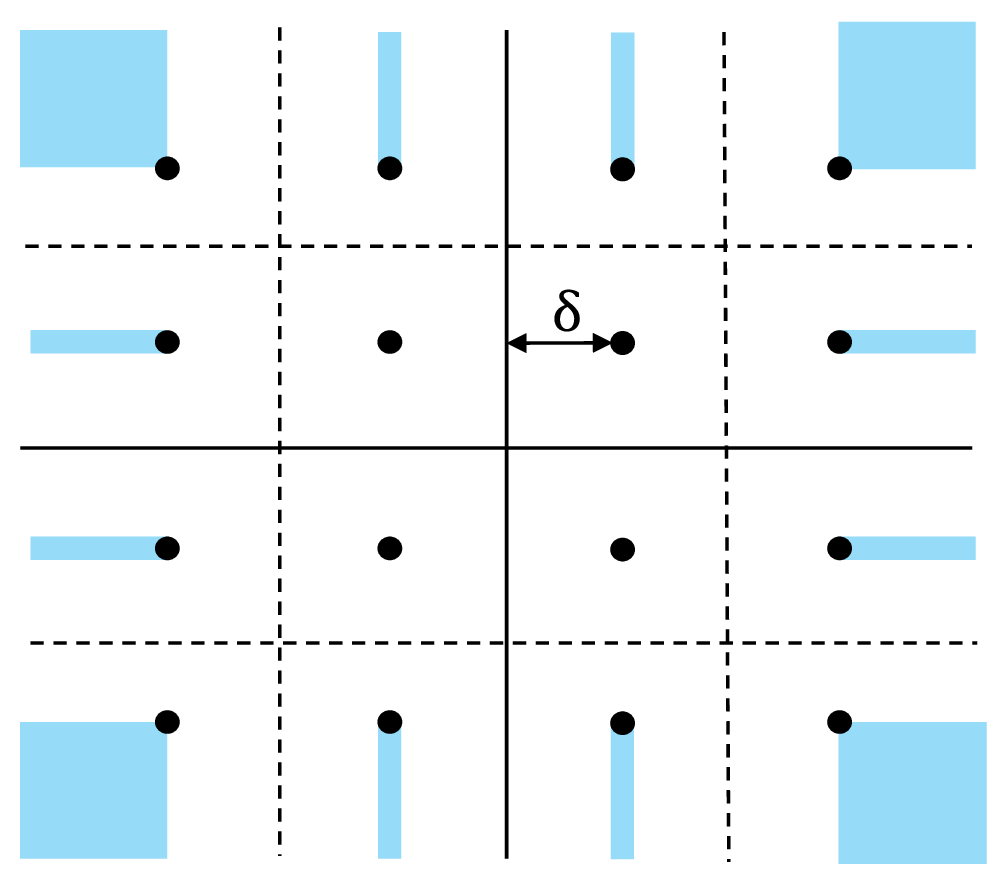}
    \caption{Illustration of conventional precoding approaches. Linear precoders tend to generate received signals near the symbols (denoted by black circles), SLP signals are constrained to lie in ``constructive'' symbol regions (blue shaded areas).}
    \label{fig:SLP_16QAM}
\end{figure}

Unlike linear precoding which treats interference as a purely negative factor, SLP is a nonlinear technique that leverages CI to push the received signals away from the symbol decision boundaries and thus enhance the detection probability. Specifically, at a time slot~$t$, the transmit signal is designed as $\mbf{x}_t = \mca{P}(\mbf{H},\mbf{s}_t)$ such that the ``safety margin'' $\delta$ is maximized, as illustrated in Fig.~\ref{fig:SLP_16QAM}. The SLP optimization for QAM signaling under a unity power constraint can be written in the following form\footnote{SLP optimizations for PSK signaling can also be found in~\cite{Junwen2023Speeding} and the references therein.}~\cite{Junwen2023Speeding}:
\begin{equation}
\begin{aligned}
    &\maximize_{\{\mbf{{x}}_t\}}  &&   {\delta}_t \\
    & \st &&  \op{sign}(s_{k,t}^\Re)\Re\{\mbf{h}_k^T\mbf{{x}}_{t}\} \unrhd {\delta}_t \op{sign}(s_{k,t}^\Re)s_{k,t}^\Re,
\;\forall k, \\
 & && \op{sign}(s_{k,t}^\Im)\Im\{\mbf{h}_k^T\mbf{{x}}_{t}\} \unrhd {\delta}_t \op{sign}(s_{k,t}^\Im)s_{k,t}^\Im,
\;\forall k,\\
    & && \|\mbf{{x}}_t\|^2 \leq 1,
    \label{eq:SLP_QAM}
\end{aligned}
\end{equation}
where ${\delta}_t$ is the minimum allowable safety margin at a time slot $t$, and the operator $\unrhd$ denotes a generalized inequality that can either be~$\geq$ or~$=$, depending on whether $s_{k,t}$ is an outer or inner constellation point, respectively (see Fig.~\ref{fig:SLP_16QAM}). Let $\mbf{\check{x}}_t$ and $\check{\delta}_t$ be the solution of~\eqref{eq:SLP_QAM} and let $\rho_t$ be the power allocated to the transmit signal signal $\mbf{x}_t$. The optimal transmit power $\rho_t$ can be obtained through the following optimization problem
\begin{equation}
\begin{aligned}
    &\maximize_{\{\rho_t\}}  &&   \delta \\
    & \st &&  \rho_t\check{\delta}_t \geq \delta \;\text{and}\; \rho_1^2 + \ldots + \rho_{\Ls}^2 \leq P,
    \label{eq:SLP_PA}
\end{aligned}
\end{equation}
which maximizes the minimum safety margin over the entire transmission block under a power constraint where $P = \rho\Ls$ is the total transmit power budget. The SLP transmit signal $\mbf{x}_t$ is then calculated as $\mbf{x}_t = \rho_t \mbf{\check{x}}_t$. Both~\eqref{eq:SLP_QAM} and~\eqref{eq:SLP_PA} are convex problems and thus can be solved efficiently. 

\subsection{Proposed CCP Framework}
Unlike conventional data-directed precoding approaches which focus on optimizing the data symbol detection performance, the proposed CCP framework aims to directly minimize the BER of the information bits $\mbf{B} = [\mbf{b}_1,\,\ldots,\,\mbf{b}_K]^T$ as illustrated in Fig.~\ref{fig:CCP}. This is motivated by the fact that, in coded systems, the ultimate goal is to recover the information bit matrix~$\mbf{B}$. Hence, CCP can be viewed as an \textit{information-directed} precoding framework. Mathematically, the CCP framework can be expressed in the following form
\begin{equation}
    \mbf{X} = \mca{P}_{\mrm{CCP}}(\mbf{H},\mbf{B})
    \label{eq:proposed_CCP_frame}
\end{equation}
where $\mca{P}_{\mrm{CCP}}(\cdot)$ is the precoder to be designed.

\begin{figure}[t!]
    \centering
    \begin{subfigure}[t]{0.42\linewidth}
        \centering
        \includegraphics[width=\linewidth]{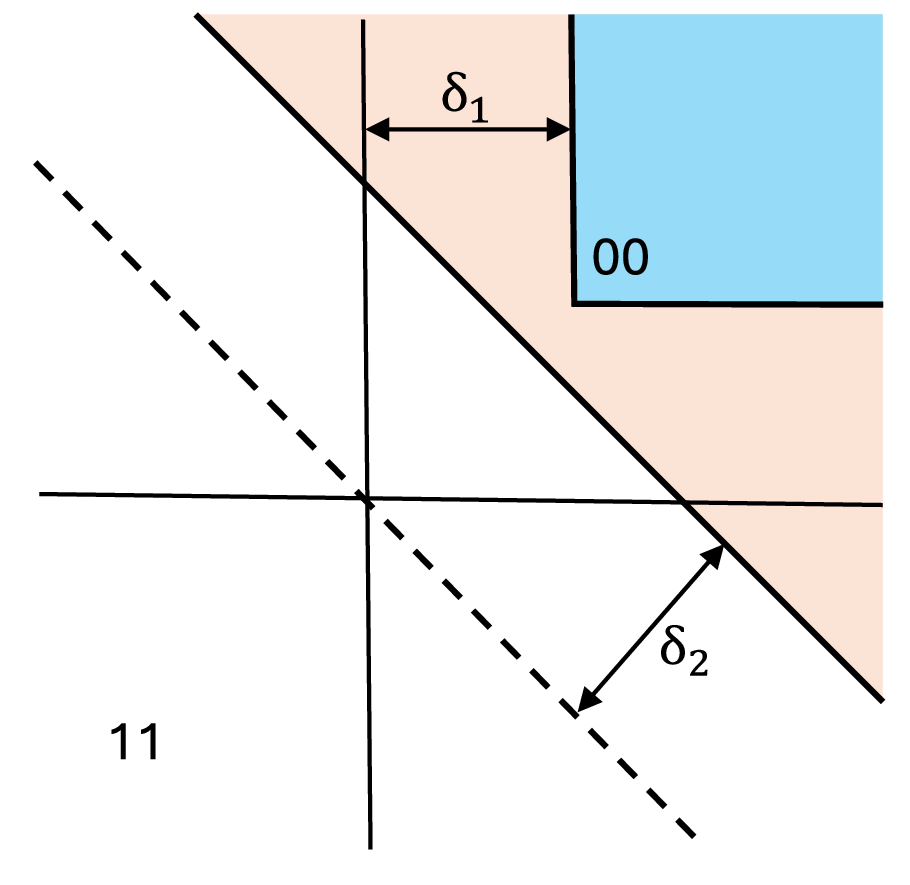}
        \caption{Constructive region comparison.}
        \label{fig:CR_rep}
    \end{subfigure}~
    \begin{subfigure}[t]{0.57\linewidth}
        \centering
        \includegraphics[width=\linewidth]{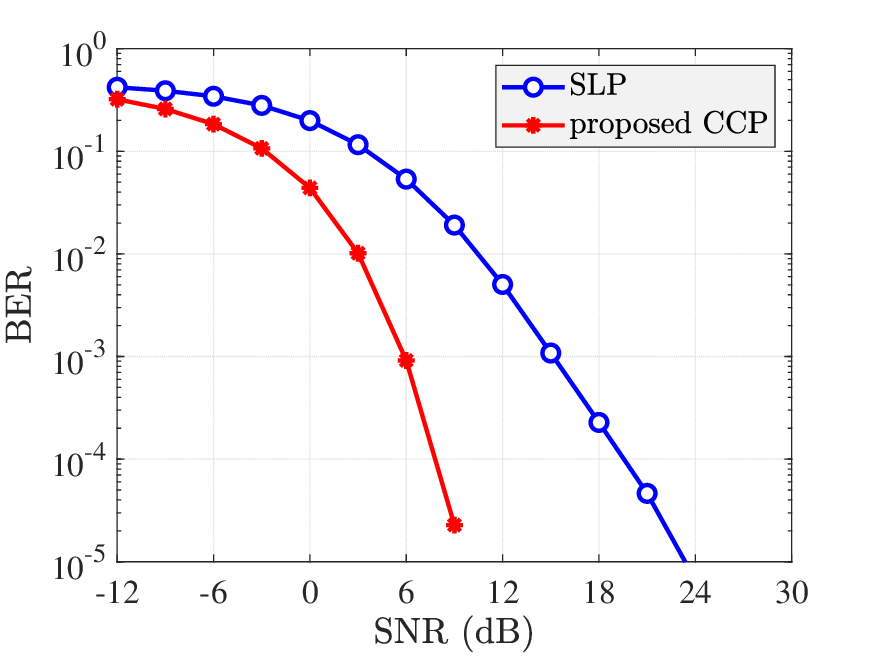}
        \caption{BER performance with a repetition code of length~3, $K = N = 4$, and 8-QAM.}
        \label{fig:BER_rep}
    \end{subfigure}
    \caption{Comparison between conventional SLP and CCP that exploits information about the repetition code.}
    \label{fig:repetition_code}
\end{figure}

The idea behind the proposed CCP framework is that exploiting information about the channel code used in the system can result in more {enhanced} signal designs. To illustrate this idea, let us consider a toy example as follows. Assume a repetition code of length~2 is employed in the system, i.e., the information bits~0 and~1 are encoded to generate two codewords~00 and~11, respectively. In conventional precoding approaches, these two codewords will be modulated to generate data symbols, which are then used as the input for the precoding process. For example, if SLP is used, the safety margin $\delta_1$ as illustrated in Fig.~\ref{fig:CR_rep} will be optimized based on the codewords. However, if we exploit the fact that there are only two possible codewords~00 and~11, we can divide the signal space into two halves and maximize the safety margin $\delta_2$. In effect, the CI region is enlarged due to the prior information regarding the restricted codeword set, which provides additional degrees of freedom for transmit signal design. In this case, the proposed optimization under a unity power constraint can be formulated as follows: 
\begin{equation}
\begin{aligned}
    &\maximize_{\{\mbf{{x}}_t\}}  &&   {\delta}_t \\
    & \st &&  \frac{1-2b_{k,t}}{\sqrt{2}}\big[\Re\{\mbf{h}_k^T\mbf{{x}}_{t}\} + \Im\{\mbf{h}_k^T\mbf{{x}}_{t}\}\big] \geq {\delta}_t,
\;\forall k
 \\
    & && \|\mbf{{x}}_t\|^2 \leq 1.
    \label{eq:SLP_repetition}
\end{aligned}
\end{equation}
In Fig.~\ref{fig:BER_rep}, we show a BER comparison for the length-3 repetition code in a system with $N = 4$ antennas serving $K = 4$ users with an 8-QAM constellation. Clearly, exploiting knowledge of the channel coding significantly enhances performance compared with the conventional data-directed SLP approach. This example demonstrates the potential of exploiting information about the channel codes when desiging the precoder. However, extending the approach in this example to more advanced channel codes is challenging due to the often large number of complex codewords involved. To accommodate advanced channel codes, we propose the CCP framework discussed below. 

Since the original objective of the CCP framework is to minimize the BER of the information bits, a direct statement of the optimization problem of interest would be to minimize the information BER under a power constraint:
\begin{equation}
\begin{aligned}
    &\minimize_{\{\mbf{X}\}}  &&   \text{BER}(\mbf{X})\\
    & \st && \|\mbf{X}\|_{F}^2 \leq P,
    \label{eq:iBER_minimization}
\end{aligned}
\end{equation}
Directly solving~\eqref{eq:iBER_minimization} is challenging due to the {complexity} of the channel coding process. To address this challenge, we propose to optimize a novel metric that \textit{minimizes the probability that the codeword received by the user has more bit errors than the error correction capacity of the employed channel code}. In effect, this optimization is designed to minimize the BER of the {\em information bits} rather than the BER of the encoded bits as in previous work. As long as the number of coded bit errors resulting from the data symbol errors is within the error-correcting capacity, they will all be corrected by the channel decoder. This is in contrast to conventional precoding approaches that attempt to eliminate errors in decoding any of the data symbols. By allowing correctable data symbol errors, the proposed CCP framework is able to exploit the extra degrees of freedom that result from channel coding for efficient transmit signal designs, thereby enhancing the overall system performance. 

In this context, the CCP framework aims to solve the following optimization problem:
\begin{equation}
\begin{aligned}
    &\maximize_{\{\mbf{X}\}}  &&   \min_k\,\mbb{P}[\|\mbf{c}_k - \mbf{\hat{c}}_k\|_0 \leq \ec \mid \mbf{X}]\\
    & \st && \|\mbf{X}\|_{F}^2 \leq P
    \label{eq:problem1}
\end{aligned}
\end{equation}
where $\|\mbf{c}_k - \mbf{\hat{c}}_k\|_0$ is the number of coded bit errors for user~$k$ and $\ec$ is the error-correcting capacity of the channel code. In particular, we maximize the worst-case probability among the users that the number of coded bit errors is no greater than the error-correcting capacity of the channel code, given a total transmit power constraint. {Solving~\eqref{eq:problem1} maximizes the probability that the channel decoder at the users can correctly recover the information bits.}

\section{Proposed CCP Design For $\ec = 1$}
\label{sec:CCP_ec1}
In this section, we propose a CCP design for channel codes with a 1-bit error-correcting capacity, i.e., $\ec = 1$. To solve the optimization problem in~\eqref{eq:problem1}, we first need to compute the probability of the event that there is no more than one coded bit error in the estimated codeword for each user $k$, which is the sum of two probabilities $\mbb{P}[\|\mbf{c}_k - \mbf{\hat{c}}_k\|_0 = 0 \mid \mbf{X}]$ and $\mbb{P}[\|\mbf{c}_k - \mbf{\hat{c}}_k\|_0 = 1 \mid \mbf{X}]$ and is given as follows:
\begin{align}
     &\mbb{P}[\|\mbf{c}_k - \mbf{\hat{c}}_k\|_0 \leq 1 \mid \mbf{X}] \notag \\
     &\quad = \mbb{P}[\|\mbf{c}_k - \mbf{\hat{c}}_k\|_0 = 0 \mid \mbf{X}] + \mbb{P}[\|\mbf{c}_k - \mbf{\hat{c}}_k\|_0 = 1 \mid \mbf{X}] \notag \\
     &\quad = \prod_{t}^{}\big(p_{k,t}^{\Re,e_0}p_{k,t}^{\Im,e_0}\big) \; + \notag \\
     &\quad \quad \, \sum_{t}^{}\Big(\big(p_{k,t}^{\Re,e_1}p_{k,t}^{\Im,e_0} + p_{k,t}^{\Re,e_0}p_{k,t}^{\Im,e_1}\big)\prod_{i\neq t}^{}\big(p_{k,i}^{\Re,e_0}p_{k,i}^{\Im,e_0}\big)\Big),
     \label{eq:Pec1}
\end{align}
where $p_{k,t}^{\Re,e_0}$ and $p_{k,t}^{\Re,e_1}$ ($p_{k,t}^{\Im,e_0}$ and $p_{k,t}^{\Im,e_1}$) respectively represent the probabilities that there are either no errors or one error in the bits corresponding to the real (imaginary) dimension of the symbols for user $k$ at time slot $t$. Eq.~\eqref{eq:Pec1}  assumes a QAM constellation, so that the error probabilities $p_{k,t}^{\Re,e_0}$ and $p_{k,t}^{\Re,e_1}$ ($p_{k,t}^{\Im,e_0}$ and $p_{k,t}^{\Im,e_1}$) can be calculated independently\footnote{{For the case of PSK constellations, the bit error probabilities cannot be independently computed for the real and imaginary parts. Instead, these probabilities are often expressed in terms of double integrals using polar coordinates~\cite{Ly2024Decision}. This complicates the transmit signal design and requires a separate study beyond the scope of this paper. Therefore, we leave the case of PSK constellations for future work.}}.

\begin{figure}[t!]
    \centering
    \begin{subfigure}[t]{0.45\linewidth}
        \centering
        \includegraphics[width=\linewidth]{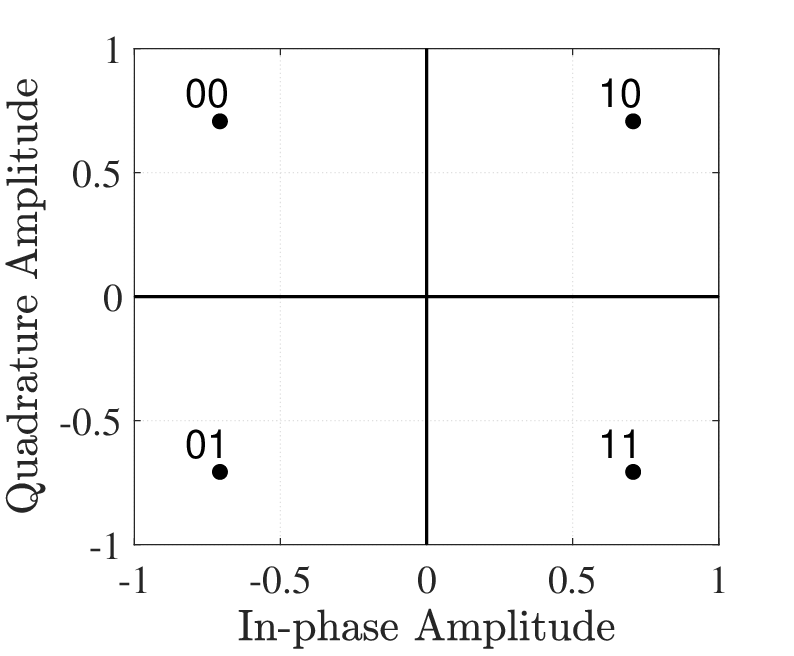}
        \caption{4-QAM}
        \label{fig:4QAM}
    \end{subfigure}~
    \begin{subfigure}[t]{0.5\linewidth}
        \centering
        \includegraphics[width=\linewidth]{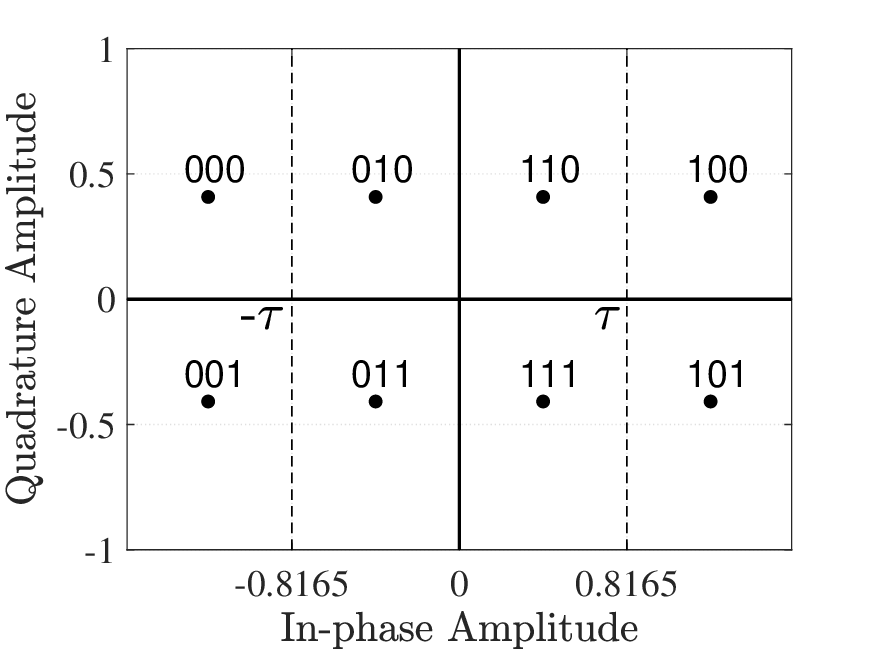}
        \caption{8-QAM}
        \label{fig:8QAM}
    \end{subfigure}
    
    \begin{subfigure}[t]{0.65\linewidth}
        \centering
        \includegraphics[width=\linewidth]{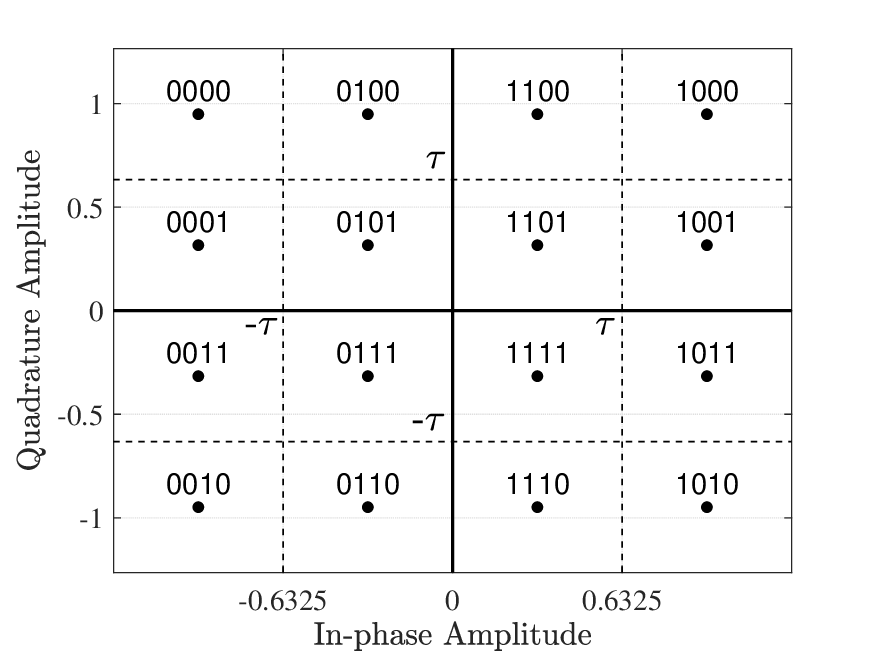}
        \caption{16-QAM}
        \label{fig:16QAM}
    \end{subfigure}
    \caption{QAM constellations with Gray coding and unit average power.}
    \label{fig:QAM_constellations}
\end{figure}

In the following, we compute the probabilities $p_{k,t}^{\Re,e_0}$, $p_{k,t}^{\Re,e_1}$, $p_{k,t}^{\Im,e_0}$, and $p_{k,t}^{\Im,e_1}$ for 4-QAM, 8-QAM, and 16-QAM constellations. We then develop a PG-based approach {to solve} problem~\eqref{eq:problem1}. It should be noted that the calculation of $p_{k,t}^{\Re,e_0}$, $p_{k,t}^{\Re,e_1}$, $p_{k,t}^{\Im,e_0}$, and $p_{k,t}^{\Im,e_1}$ specifically depends on how the bit sequences are mapped onto the constellation points. We will consider the common case of Gray coding, as illustrated in Fig.~\ref{fig:QAM_constellations}.

\subsection{Coded Bit Error Probability}
For convenience in later derivations, we write the noiseless received signal for user $k$ at time $t$ using the following real-valued formulation:
\begin{align*}
    \begin{bmatrix}z_{k,t}^\Re \\z_{k,t}^\Im\end{bmatrix} = \begin{bmatrix}
        \Re\{\mbf{{h}}_k\}^T & -\Im\{\mbf{{h}}_k^T\}
        \\
        \Im\{\mbf{{h}}_k\}^T & \Re\{\mbf{{h}}_k^T\}
    \end{bmatrix} \begin{bmatrix}\Re\{\mbf{x}_t\}\\\Im\{\mbf{x}_t\}\end{bmatrix}.
\end{align*}
The corresponding noisy received signal can then be written~as
\begin{align*}
        \Re\{y_{k,t}\} &= z_{k,t}^\Re + \Re\{n_{k,t}\},\\
        \Im\{y_{k,t}\} &= z_{k,t}^\Im + \Im\{n_{k,t}\}.
\end{align*}

\subsubsection{4-QAM} Given the 4-QAM constellation in Fig.~\ref{fig:4QAM}, the real part of the received signal ${y}_{k,t}$ determines the estimated bit $\hat{c}_{k,2(t-1)+1}$. The probabilities that $\hat{c}_{k,2(t-1)+1}$ is estimated as 0 or 1 are given by
\begin{align}
    \mbb{P}[\hat{c}_{k,2(t-1)+1} = 1 \mid \mbf{X}] &= \mbb{P}[\Re\{{y}_{k,t}\} \geq 0] = \Phi(\sqrt{2}z_{k,t}^\Re/\sigma), \notag\\
    \mbb{P}[\hat{c}_{k,2(t-1)+1} = 0 \mid \mbf{X}] &= \mbb{P}[\Re\{{y}_{k,t}\} < 0] = \Phi(-\sqrt{2}z_{k,t}^\Re/\sigma). \notag
\end{align}
Therefore, we can write $p_{k,t}^{\Re,e_0}$ in the following compact form:
\begin{align}
    p_{k,t}^{\Re,e_0} &= \begin{cases}
        \mbb{P}[\hat{c}_{k,2(t-1)+1} = 1 \mid \mbf{X}] \;\text{if}\; c_{k,2(t-1)+1} = 1\\
        \mbb{P}[\hat{c}_{k,2(t-1)+1} = 0 \mid \mbf{X}] \;\text{if}\; c_{k,2(t-1)+1} = 0
    \end{cases} \notag \\
    &= \Phi(\sqrt{2}(2c_{k,2(t-1)+1}-1) z_{k,t}^\Re/\sigma).
\end{align}
The probability $p_{k,t}^{\Re,e_1}$ is given by
\begin{align}
    p_{k,t}^{\Re,e_1} &=\Phi(-\sqrt{2}(2c_{k,2(t-1)+1}-1) z_{k,t}^\Re/\sigma)\notag \\
    &= 1- p_{k,t}^{\Re,e_0}.
\end{align}
Similarly, the imaginary part of the received signal ${y}_{k,t}$ determines the estimated bit $\hat{c}_{k,2(t-1)+2}$. We then have
\begin{align}
    p_{k,t}^{\Im,e_0} &= \Phi(-\sqrt{2}(2c_{k,2(t-1)+2}-1) z_{k,t}^\Im/\sigma), \\
    p_{k,t}^{\Im,e_1} &= 1 - p_{k,t}^{\Im,e_0}.
\end{align}
    
\subsubsection{8-QAM}
Unlike the 4-QAM constellation where only one bit is modulated onto the real dimension, the real part of the 8-QAM constellation (see Fig.~\ref{fig:8QAM}) carries two bits. Thus, the real part of ${y}_{k,t}$ determines $\{\hat{c}_{k,3(t-1)+1},\hat{c}_{k,3(t-1)+2}\}$. There are four possibilities for $\{c_{k,3(t-1)+1}, c_{k,3(t-1)+2}\}$, and the probabilities $p_{k,t}^{\Re,e_0}$ and $p_{k,t}^{\Re,e_1}$ are calcuated below for each case. The variable $\tau$ in the following derivations is the decision threshold employed by the users for data detection.

\textbf{Case 1:} $\{c_{k,3(t-1)+1}, c_{k,3(t-1)+2}\} = \{0,0\}$. The probability of no bit errors is
\begin{align}
    p_{k,t}^{\Re,e_0} &=  \mbb{P}[\{\hat{c}_{k,3(t-1)+1}, \hat{c}_{k,3(t-1)+2}\} = \{0,0\} \mid \mbf{X}]\notag\\
    &=\mbb{P}[\Re\{{y}_{k,t}\} \leq -\tau \mid \mbf{X}] = \Phi(\sqrt{2}(-\tau-z_{k,t}^\Re)/\sigma),
\end{align}
and the probability of one bit error is
\begin{align}
    p_{k,t}^{\Re,e_1} &= \mbb{P}[\{\hat{c}_{k,3(t-1)+1}, \hat{c}_{k,3(t-1)+2}\} = \{0,1\} \mid \mbf{X}] \; + \notag \\
    &\quad\; \mbb{P}[\{\hat{c}_{k,3(t-1)+1}, \hat{c}_{k,3(t-1)+2}\} = \{1,0\} \mid \mbf{X}] \notag \\
    &= \mbb{P}[-\tau \leq \Re\{{y}_{k,t}\} \leq 0 \mid \mbf{X}] + \mbb{P}[\Re\{{y}_{k,t}\} \geq \tau \mid \mbf{X}] \notag \\
    &= \Phi(-\sqrt{2}z_{k,t}^\Re/\sigma) - \Phi(\sqrt{2}(-\tau-z_{k,t}^\Re)/\sigma) \; + \notag \\
    &\quad \; \Phi(\sqrt{2}(-\tau+z_{k,t}^\Re)/\sigma).
\end{align}
We can similarly obtain the probabilities of no bit errors $p_{k,t}^{\Re,e_0}$ and one bit error $p_{k,t}^{\Re,e_1}$ for the other cases.

\textbf{Case 2:} $\{c_{k,3(t-1)+1}, c_{k,3(t-1)+2}\} = \{0,1\}$.
\begin{align}
    p_{k,t}^{\Re,e_0} &=  \Phi(-\sqrt{2}z_{k,t}^\Re/\sigma) - \Phi(\sqrt{2}(-\tau-z_{k,t}^\Re)/\sigma),\\
    p_{k,t}^{\Re,e_1} &=  \Phi(\sqrt{2}(-\tau-z_{k,t}^\Re)/\sigma) \; + \notag \\
    &\quad \; \Phi(\sqrt{2}(\tau -z_{k,t}^\Re)/\sigma) - \Phi(\sqrt{2}(-z_{k,t}^\Re)/\sigma).
\end{align}

\textbf{Case 3:} $\{c_{k,3(t-1)+1}, c_{k,3(t-1)+2}\} = \{1,0\}$.
\begin{align}
    p_{k,t}^{\Re,e_0} &=  \Phi(\sqrt{2}(-\tau+z_{k,t}^\Re)/\sigma), \\
    p_{k,t}^{\Re,e_1} &=  \Phi(\sqrt{2}(-\tau-z_{k,t}^\Re)/\sigma) \; + \notag \\
    &\quad \; \Phi(\sqrt{2}(\tau -z_{k,t}^\Re)/\sigma) - \Phi(\sqrt{2}(-z_{k,t}^\Re)/\sigma).
\end{align}

\textbf{Case 4:} $\{c_{k,3(t-1)+1}, c_{k,3(t-1)+2}\} = \{1,1\}$.
\begin{align}
    p_{k,t}^{\Re,e_0} &=\Phi(\sqrt{2}(\tau -z_{k,t}^\Re)/\sigma) - \Phi(\sqrt{2}(-z_{k,t}^\Re)/\sigma), \\
    p_{k,t}^{\Re,e_1} &= \Phi(\sqrt{2}(-z_{k,t}^\Re)/\sigma) - \Phi(\sqrt{2}(-\tau-z_{k,t}^\Re)/\sigma) \; + \notag \\
    &\quad \; \Phi(\sqrt{2}(-\tau+z_{k,t}^\Re)/\sigma).
\end{align}

Unlike the real dimension which conveys two bits, the imaginary part of the received signal ${y}_{k,t}$ determines only one estimated bit, which is $\hat{c}_{k,3(t-1)+3}$. Therefore, the probabilities $p_{k,t}^{\Im,e_0}$ and $p_{k,t}^{\Im,e_1}$ can be written in compact forms like the case of 4-QAM as follows:
\begin{align}
    p_{k,t}^{\Im,e_0} &=  \Phi(-\sqrt{2}(2c_{k,3(t-1)+3}-1) z_{k,t}^\Im/\sigma)\\
    p_{k,t}^{\Im,e_1} &=  1 - p_{k,t}^{\Im,e_0}.
\end{align}

\subsubsection{16-QAM}
For 16-QAM, the real and imaginary parts of the received signal ${y}_{k,t}$ determine the estimated two-bit sequences $\{\hat{c}_{k,4(t-1)+1},\hat{c}_{k,4(t-1)+2}\}$ and $\{\hat{c}_{k,4(t-1)+3},\hat{c}_{k,4(t-1)+4}\}$, respectively. Therefore, all of the probabilities $p_{k,t}^{\Re,e_0}$, $p_{k,t}^{\Re,e_1}$, $p_{k,t}^{\Im,e_0}$ and $p_{k,t}^{\Im,e_1}$ must be computed for the four different cases considered for 8-QAM. As the derivations are similar, we omit them here for brevity.

\begin{figure*}[t!]
    \begin{align}
     \nabla \mbb{P}[\|\mbf{c}_k - \mbf{\hat{c}}_k\|_0 \leq 1] &= \sum_{t}^{}\big(\nabla p_{k,t}^{\Re,e_0}p_{k,t}^{\Im,e_0} + p_{k,t}^{\Re,e_0}\nabla p_{k,t}^{\Im,e_0}\big)\prod_{i\neq t}^{}\big(p_{k,i}^{\Re,e_0}p_{k,i}^{\Im,e_0}\big) \; + \notag \\
     &\quad\,\sum_{t}^{}\big(\nabla p_{k,t}^{\Re,e_1}p_{k,t}^{\Im,e_0} + p_{k,t}^{\Re,e_1}\nabla p_{k,t}^{\Im,e_0} + \nabla p_{k,t}^{\Re,e_0}p_{k,t}^{\Im,e_1} + p_{k,t}^{\Re,e_0}\nabla p_{k,t}^{\Im,e_1}\big)\prod_{i\neq t}^{}\big(p_{k,i}^{\Re,e_0}p_{k,i}^{\Im,e_0}\big)\; + \notag \\
     &\quad\,\sum_{t}^{}\big(p_{k,t}^{\Re,e_1}p_{k,t}^{\Im,e_0} + p_{k,t}^{\Re,e_0}p_{k,t}^{\Im,e_1}\big)\sum_{i\neq t}^{}\big(\nabla p_{k,i}^{\Re,e_0}p_{k,i}^{\Im,e_0} + p_{k,i}^{\Re,e_0}\nabla p_{k,i}^{\Im,e_0}\big)\prod_{j\neq i,t}^{}\big(p_{k,j}^{\Re,e_0}p_{k,j}^{\Im,e_0}\big).
     \label{eq:grad_Pe1}
\end{align}
\end{figure*}

\subsection{Proposed PG-based CCP Design}
We have computed the probability $\mbb{P}[\|\mbf{c}_k - \mbf{\hat{c}}_k\|_0 \leq 1 \mid \mbf{X}]$, which constitutes the objective function in problem~\eqref{eq:problem1}. It can be seen that this is a complex non-convex and non-linear objective function that is challenging to solve. In this section, we develop a PG-based approach to find a local minimum for problem~\eqref{eq:problem1}.

\begin{algorithm}[t!]
    \small
    \KwIn{$\mbf{H}$, $\mathbf{\tilde{x}}^{(0)}$, $\alpha_0$, $\eta$.}
    \textbf{Initialization:} $\ell = 0$, $\kappa = 0$, $\alpha = \alpha_0$, $p_{\mrm{min}}^\star = 0$\;
    \While{$\mrm{stop} = \mrm{false}$} 
    {
        Obtain $\hat{k}$ by $\hat{k} = \argmin_{k}\, \mbb{P}[\|\mbf{c}_k - \mbf{\hat{c}}_k\|_0 \leq 1 \mid \mbf{\tilde{x}}^{(\ell)}]$\label{algo:khat}\;
        Set $p_{\mrm{min}} = \mbb{P}[\|\mbf{c}_{\hat{k}} - \mbf{\hat{c}}_{\hat{k}}\|_0 \leq 1 \mid \mbf{\tilde{x}}^{(\ell)}]$\label{algo:pmin}\;
        \eIf{$p_{\mrm{min}} > p_{\mrm{min}}^\star$}{
            Update $\mbf{\tilde{x}}^\star = \mbf{\tilde{x}}^{(\ell)}$ and $p_{\mrm{min}}^\star = p_{\mrm{min}}$\label{algo:bestsol}\;
            Set $\kappa = 0$\;
        }{
            Set $\kappa = \kappa + 1$\;
            \If{$\kappa = \kappa_{\mrm{max}}$}{
                Set $\alpha = \eta \alpha$\label{algo:updatestepsize}\;
                \eIf{$\alpha \leq \alpha_{\mrm{min}}$}{
                Set $\mrm{stop} = \mrm{true}$\;
                }{
                Set $\kappa = 0$ and $\mbf{\tilde{x}}^{(\ell)} = \mbf{\tilde{x}}^\star$\;}
            }
        }
        \eIf{$\ell = \ell_{\mrm{max}}$}{
            $\mrm{stop} = \mrm{true}$\;
        }{
            $\mbf{\tilde{x}}^{(\ell+1)} = \Pi(\mbf{\tilde{x}}^{(\ell)} + \alpha \nabla \mbb{P}[\|\mbf{c}_{\hat{k}} - \mbf{\hat{c}}_{\hat{k}}\|_0 \leq 1 \mid \mbf{\tilde{x}}^{(\ell)}])$\label{algo:updatex}\;
            Set $\ell = \ell + 1$\;
        }
    }
    \Return{$\mbf{\tilde{x}}^\star$}\;
    \caption{Proposed PG-based CCP for $\ec = 1$ and 4-QAM.}
    \label{algo:gradCCP}
\end{algorithm}

Let
\begin{equation*}
    \mbf{\tilde{x}}_t = \begin{bmatrix}\Re\{\mbf{x}_t\}\\\Im\{\mbf{x}_t\}\end{bmatrix}\;\text{and}\;
    \mbf{\tilde{x}} = \begin{bmatrix}[0.5]\mbf{\tilde{x}}_1\\ \vdots \\ \mbf{\tilde{x}}_{\Ls}\end{bmatrix}
\end{equation*}
be the real-valued representations of the transmit signal at time slot $t$ and the entire block, respectively. To solve for $\mbf{\tilde{x}}$, we employ the PG-based method as follows: 
\begin{equation}
    \mbf{\tilde{x}}^{(\ell+1)} = \Pi(\mbf{\tilde{x}}^{(\ell)} + \alpha \times \nabla \mbb{P}[\|\mbf{c}_{\hat{k}} - \mbf{\hat{c}}_{\hat{k}}\|_0 \leq 1 \mid \mbf{\tilde{x}}^{(\ell)}]),
\end{equation}
where $\ell$ is the iteration index, $\alpha$ is a step size, $\hat{k}$ is given as
\begin{equation}
    \hat{k} = \argmin_{k}\, \mbb{P}[\|\mbf{c}_k - \mbf{\hat{c}}_k\|_0 \leq 1 \mid \mbf{\tilde{x}}^{(\ell)}],
    \label{eq:khat}
\end{equation}
and $\Pi(\cdot)$ is a projection function to guarantee the transmit power constraint, defined as $\Pi(\mbf{a}) = \mbf{a}$ if $\|\mbf{a}\|_2^2 \leq P$ and $\Pi(\mbf{a}) = \sqrt{P}\,\mbf{a}/\|\mbf{a}\|_2$, otherwise.
The gradient $\nabla \mbb{P}[\|\mbf{c}_k - \mbf{\hat{c}}_k\|_0 \leq 1 \mid \mbf{\tilde{x}}]$ is given in~\eqref{eq:grad_Pe1}, which not only depends on the probabilities $ p_{k,t}^{\Re,e_0}$, $ p_{k,t}^{\Re,e_1}$, $ p_{k,t}^{\Im,e_0}$, and $p_{k,t}^{\Im,e_1}$, but also their gradients $\nabla p_{k,t}^{\Re,e_0}$, $\nabla p_{k,t}^{\Re,e_1}$, $\nabla p_{k,t}^{\Im,e_0}$, and $\nabla p_{k,t}^{\Im,e_1}$ as well. To this end, let
\begin{align*}
    [\mbf{g}_{k,1},\,\ldots,\,\mbf{g}_{k,2\Ls}]^T = \mbf{I}_{\Ls}\otimes \begin{bmatrix}
        \Re\{\mbf{{h}}_k^T\} & -\Im\{\mbf{{h}}_k^T\}
        \\
        \Im\{\mbf{{h}}_k^T\} & \Re\{\mbf{{h}}_k^T\}
    \end{bmatrix},
\end{align*}
so the gradients for the case of a 4-QAM constellation can be obtained as follows:
\begin{align}
    \nabla_{\mbf{\tilde{x}}}\, p_{k,t}^{\Re,e_0} &= \frac{2c_{k,2(t-1)+1}-1}{\sigma\sqrt{\pi}}\mbf{g}_{k,2(t-1)+1}\,e^{-(z_{k,t}^\Re/\sigma)^2}, \\
    \nabla_{\mbf{\tilde{x}}}\, p_{k,t}^{\Re,e_1} &= -\nabla_{\mbf{\tilde{x}}}\, p_{k,t}^{\Re,e_0},\\
    \nabla_{\mbf{\tilde{x}}}\, p_{k,t}^{\Im,e_0} &= -\frac{2c_{k,2(t-1)+2}-1}{\sigma\sqrt{\pi}}\mbf{g}_{k,2(t-1)+2}\,e^{-(z_{k,t}^\Im/\sigma)^2}, \label{eq:grad_pIm0_4QAM} \\
    \nabla_{\mbf{\tilde{x}}}\, p_{k,t}^{\Im,e_1} &= -\nabla_{\mbf{\tilde{x}}}\, p_{k,t}^{\Im,e_0}. \label{eq:grad_pIm1_4QAM}
\end{align}

A detailed description of the proposed PG-based CCP design for $\ec = 1$ and 4-QAM is given in Algorithm~\ref{algo:gradCCP}. Given a transmit signal $\mbf{\tilde{x}}^{(\ell)}$, we first find the user index associated with the lowest probability in line~\ref{algo:khat} and the corresponding value of the objective function $p_{\mrm{min}}$ in line~\ref{algo:pmin}. Based on this, we keep track of the best solution in line~\ref{algo:bestsol} and update the solution $\mbf{\tilde{x}}^{(\ell+1)}$ in line~\ref{algo:updatex}. Note that the step size $\alpha$ is reduced by a factor $\eta < 1$ in line~\ref{algo:updatestepsize} if the value of the objective does not improve after $\kappa_{\mrm{max}}$ consecutive iterations.

For the case of 4-QAM, there is only a single decision threshold at zero and only the transmit signal $\mbf{\tilde{x}}$ needs to be optimized. However, for 8-QAM and 16-QAM, the decision threshold $\tau$ must be determined together with $\mbf{\tilde{x}}$. To this end, we define the augmented vector 
\begin{equation}
    \mbf{\bar{x}} = \begin{bmatrix}
        \tau \\ 
        \mbf{\tilde{x}}
    \end{bmatrix}
\end{equation}
and calculate the gradients $\nabla p_{k,t}^{\Re,e_0}$, $\nabla p_{k,t}^{\Re,e_1}$, $\nabla p_{k,t}^{\Im,e_0}$, and $\nabla p_{k,t}^{\Im,e_1}$ for 8-QAM and 16-QAM with respect to (w.r.t) $\mbf{\bar{x}}$ instead of only $\mbf{\tilde{x}}$ as in the case of 4-QAM. We provide details for calculating these gradients in Appendix~\ref{appendix_1}. The proposed CCP design for 8-QAM and 16-QAM with $\ec = 1$ can also be implemented by Algorithm~\ref{algo:gradCCP}, with the exception that $\mbf{\tilde{x}}$ is replaced by~$\mbf{\bar{x}}$.

{The computational complexity of Algorithm~\ref{algo:gradCCP} is $\mca{O}(NK\Ls)$ per iteration, due mainly to the computation of the objective function and the corresponding gradients. In our investigations, we observed that although the sequence of objective function values obtained by Algorithm~\ref{algo:gradCCP} does not monotonically increase, it converges to at least a local solution. The initialization also affects the performance of the proposed CCP approach, and we will demonstrate later that initializing the algorithm with the conventional SLP approach performs well.}

\section{Proposed CCP Design for $\ec > 1$}
\label{sec:CCP_ecg1}
In the previous section, we considered the case where the error-correcting capacity of the channel code is only one bit ($\ec = 1$). In this section, we consider channel codes with a stronger error-correcting capacity of more than one bit, i.e., $\ec > 1$. In this case, the probability that the number of bit errors $\|\mbf{c}_k - \mbf{\hat{c}}_k\|_0$ is less than or equal to $\ec$ is given by
\begin{align*}
    \mbb{P}[\|\mbf{c}_k - \mbf{\hat{c}}_k\|_0 \leq \ec \mid \mbf{X}] &= \sum_{i=0}^{\ec} \mbb{P}[\|\mbf{c}_k - \mbf{\hat{c}}_k\|_0 = i]\notag \\
    &= \sum_{i=0}^{\ec} \sum_{j=1}^{\binom{\Lc}{i}} \mbb{P}[f_{\mrm{nz}}(\mbf{c}_k - \hat{\mbf{c}}_k) = \mca{L}_{i,j}],
\end{align*}
where $f_{\mrm{nz}}(\mbf{a})$ denotes the set of indices corresponding to the non-zero elements of $\mbf{a}$, the binomial coefficient $\binom{\Lc}{i}$ is the number of possibilities for which there are $i$ bit errors in an $\Lc$-bit codeword, and $\mca{L}_{i,j}$ denotes the set of the $i$ bit error indices corresponding to the $j$-th possibility.

Directly optimizing the above probability is challenging since the binomial coefficient $\binom{\Lc}{i}$ scales quickly in terms of $\Lc$. To simplify the problem, we propose to divide the entire block into $\Nsb$ sub-blocks as $\mbf{X} = [\mbf{X}_1,\,\ldots,\,\mbf{X}_{\Nsb}]$ and design the transmit signal in these sub-blocks separately. Specifically, within each sub-block $i$, we design the transmit signal $\mbf{X}_i$ so that it maximizes the probability there is no more than one bit error in the sub-block. This means we solve the following optimization problem for the $i$-th sub-block:
\begin{equation}
\begin{aligned}
    &\maximize_{\{\mbf{X}_i\}} && \min_k\,\mbb{P}[\|\mbf{c}_{k,i} - \mbf{\hat{c}}_{k,i}\|_0 \leq 1 \mid \mbf{X}_i]\\
    & \st && \|\mbf{X}_i\|_{F}^2 \leq P/\Nsb,
    \label{eq:problem2}
\end{aligned}
\end{equation}
where $\mbf{c}_{k,i}$ and $\mbf{\hat{c}}_{k,i}$ are the vector of coded bits within the $i$-th sub-block of user $k$ and the corresponding detected bits, respectively. Clearly, choosing an appropriate number of sub-blocks is critical since it significantly affects the precoding performance. It is obvious that a lower bound for the number of sub-blocks $\Nsb$ would be the value of $\ec$. We will examine the effect of $\Nsb$ later in the numerical result section and show that there is an optimal value that gives the best performance.

Using the above approach, we can more efficiently address the precoding problem for the case $\ec > 1$, solving the simpler optimization problem outlined in the previous section with $\ec=1$ for each sub-block. However, for 8-QAM and 16-QAM, this approach would require the design of $\Nsb$ different values of the decision threshold $\tau$, one for each of the sub-blocks. This in turn increases the feedback overhead from the BS to the users since $\tau$ must be known at the users. An alternative approach is to design a single decision threshold value $\tau$ for the entire block while still separately designing $\mbf{X}_i$ for each sub-block $i$ using~\eqref{eq:problem2}. We propose an algorithm for this alternative approach and provide a detailed description in Algorithm~\ref{algo:gradCCP_tau_design}. Specifically, Algorithm~\ref{algo:gradCCP_tau_design} alternately optimizes the decision threshold $\tau$ and the transmit signal matrices $\mbf{X}_i$. For a given value of $\tau$, the matrices $\mbf{X}_i \, \forall i$ are found using the approach of Section~\ref{sec:CCP_ec1} (see line~\ref{algo2:updateXi} of Algorithm~\ref{algo:gradCCP_tau_design}). Then, with fixed $\mbf{X}_i$, we either increase $\tau$ by a factor $\zeta > 1$ on line~\ref{algo2:increase_tau}, or we reduce $\tau$ by a factor $\xi < 1$ on line~\ref{algo2:reduce_tau} if the value of the objective function is respectively larger or smaller than that for the best solution found previously. This $\tau$-update strategy is based on the observation that performance improvement may be achieved from larger symbol regions as $\tau$ increases. Therefore, in our design, we increase the value of $\tau$ as long as we obtain a larger value of the objective function. The algorithm stops at line~\ref{algo2:stop} when $\tau$ can no longer be increased for a larger objective function value.

The computational complexity of line~\ref{algo2:updateXi} in Algorithm~\ref{algo:gradCCP_tau_design} is $\mca{O}(I_1NK\Ls/\Nsb)$ where $I_1$ is the number of iterations for Algorithm~\ref{algo:gradCCP} to converge. However, line~\ref{algo2:updateXi} is implemented $\Nsb$ times within each iteration of the \textbf{while} loop. Since the computational complexity of Algorithm~\ref{algo:gradCCP_tau_design} is mainly from implementing the approach in line~\ref{algo2:updateXi}, the total computational complexity of Algorithm~\ref{algo:gradCCP_tau_design} is $\mca{O}(I_2I_1NK\Ls)$ where $I_2$ is the number of iterations from the \textbf{while} loop of Algorithm~\ref{algo:gradCCP_tau_design}. We also have the similar conclusion that  Algorithm~\ref{algo:gradCCP_tau_design} converges to at least a local optimum.

\begin{algorithm}[t!]
    \small
    \KwIn{$\mbf{H}$, $\mbf{X}^{(0)}_i$, $\tau_0$, $\xi < 1$, and $\zeta > 1$.}
    Set $\mbf{X}^\star_i = \mbf{X}^{(0)}_i$, $\tau^\star = \tau_0$, $\tau = \zeta\tau$\;
    Set $p_{\mrm{min}}^\star =  \min_{k,i}\,\mbb{P}[\|\mbf{c}_{k,i} - \mbf{\hat{c}}_{k,i}\|_0 \leq 1 \mid \mbf{X}^\star_i, \tau^\star]$\;
    \While{$\mrm{stop} = \mrm{false}$} 
    {
        \For{$i=1,\,\ldots,\,\Nsb$}{
            Implement the approach of Section~\ref{sec:CCP_ec1} using the current $\tau$ to design $\mbf{X}_i$\label{algo2:updateXi}\;
        }
        Obtain $p_{\mrm{min}} = \min_{k,i}\,\mbb{P}[\|\mbf{c}_{k,i} - \mbf{\hat{c}}_{k,i}\|_0 \leq 1 \mid \mbf{X}_i, \tau]$\;
        \eIf{$p_{\mrm{min}} \leq p_{\mrm{min}}^\star$}{
            Update $\tau = \xi \tau$\label{algo2:reduce_tau}\;
            \If{$\tau \leq \tau^\star$}{
                $\mrm{stop} = \mrm{true}$\label{algo2:stop}\;
            }
        }{
            Update $p_{\mrm{min}}^\star = p_{\mrm{min}}$, $\mbf{X}^\star_i = \mbf{X}_i$, and $\tau^\star = \tau$\;
            Set $\tau = \zeta\tau$\label{algo2:increase_tau}\;
        }
        
    }
    \Return{$\mbf{X}^\star_i$, $\tau^\star$}\;
    \caption{Proposed PG-based CCP for joint $\mbf{X}_i$ and $\tau$ design with $\ec > 1$.}
    \label{algo:gradCCP_tau_design}
\end{algorithm}

\section{Robust CCP Design}
\label{sec:robust_CCP}
Thus far, we have developed the CCP approach under the assumption of perfect CSI. In this section, we consider imperfect CSI and propose a robust CCP design, taking into account the effect of both channel estimation error at the BS and noise at the users. We assume a channel estimate model $\mbf{\hat{H}} = \mbf{{H}} + \mbf{E}$, where $\mbf{E}$ represents a matrix of estimation errors whose elements are independent and identically distributed (i.i.d.) as $\mca{CN}(0,\varepsilon^2)$. Define
\begin{align}
    \begin{bmatrix}\hat{z}_{k,t}^\Re \\\hat{z}_{k,t}^\Im\end{bmatrix} = \begin{bmatrix}
        \Re\{\mbf{\hat{h}}_k^T\} & -\Im\{\mbf{\hat{h}}_k^T\}
        \\
        \Im\{\mbf{\hat{h}}_k^T\} & \Re\{\mbf{\hat{h}}_k^T\}
    \end{bmatrix} \begin{bmatrix}\Re\{\mbf{x}_t\}\\\Im\{\mbf{x}_t\}\end{bmatrix}
\end{align}
so that we can write
\begin{align*}
    z_{k,t}^\Re = \hat{z}_{k,t}^\Re + w_{k,t}^\Re\\
    z_{k,t}^\Im = \hat{z}_{k,t}^\Im + w_{k,t}^\Im,
\end{align*}
where $w_{k,t}^\Re \sim \mca{N}(0,\varepsilon^2\|\mbf{\tilde{x}}_{t}\|_2^2/2)$ and $w_{k,t}^\Im \sim \mca{N}(0,\varepsilon^2\|\mbf{\tilde{x}}_t\|_2^2/2)$. It can be seen that the design of $\mbf{\tilde{x}}_t$ contributes to the impact of the channel estimation error since the variance of $w_{k,t}$ depends on $\mbf{x}_t$.

\begin{figure}[t!]
    \centering
    \begin{subfigure}[t]{0.9\linewidth}
        \centering
        \includegraphics[width=\linewidth]{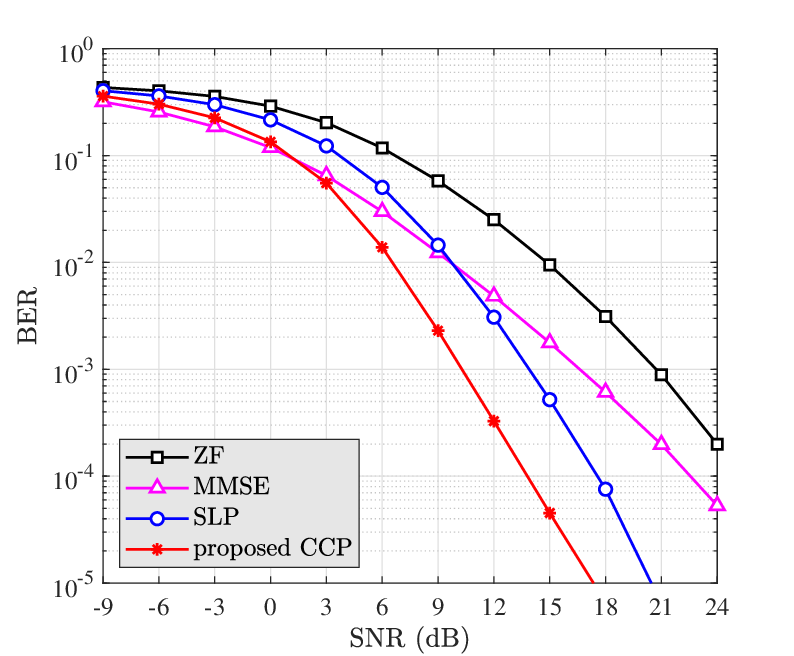}
        \caption{BER performance.}
        \label{fig:BER_vs_SNR_Hamming74_4K_4N}
    \end{subfigure}

    \begin{subfigure}[t]{0.9\linewidth}
        \centering
        \includegraphics[width=\linewidth]{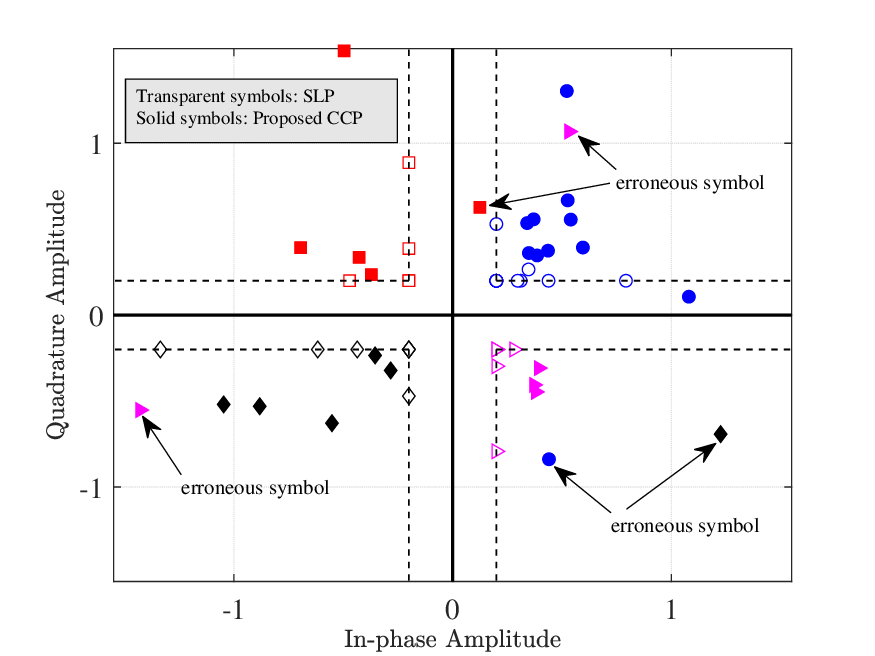}
        \caption{Noiseless received signals.}
        \label{fig:noiseless_rx_signal}
    \end{subfigure}
    \caption{BER performance and noiseless received signals for $\ec = 1$ (Hamming [7,4]), $K = N = 4$, and 4-QAM constellation.}
    \label{fig:performance_Hamming74_4K_4N}
\end{figure}

For computation of the error probabilities and the corresponding gradients, we need to replace $z_{k,t}^\Re$, $z_{k,t}^\Im$, and $\sigma$ by $\hat{z}_{k,t}^\Re$, $\hat{z}_{k,t}^\Im$, and $\varsigma_t$ respectively, where $\varsigma^2_t = \sigma^2 + \varepsilon^2\|\mbf{\tilde{x}}_t\|_2^2$ represents the variance of the effective error. For example, in case of 4-QAM, the probabilities $p_{k,t}^{\Re,e_0}$, $p_{k,t}^{\Re,e_1}$, $p_{k,t}^{\Im,e_0}$, and $p_{k,t}^{\Im,e_1}$ are given as follows:
\begin{align}
    p_{k,t}^{\Re,e_0} &= \Phi\left(\sqrt{2}(2c_{k,2(t-1)+1}-1) \hat{z}_{k,t}^\Re/\varsigma_t\right),\notag \\
    p_{k,t}^{\Re,e_1} &= 1 - p_{k,t}^{\Re,e_0}, \notag \\
    p_{k,t}^{\Im,e_0} &= \Phi\left(-\sqrt{2}(2c_{k,2(t-1)+2}-1) \hat{z}_{k,t}^\Im/\varsigma_t\right),\notag \\
    p_{k,t}^{\Im,e_1} &= 1 - p_{k,t}^{\Im,e_0}, \notag
\end{align}
while their gradients w.r.t $\mbf{\tilde{x}}$ are given by
\begin{align}
    &\nabla_{\mbf{\tilde{x}}}\, p_{k,t}^{\Re,e_0} = e^{-(\hat{z}_{k,t}^\Re)^2/\varsigma_t^2} \; \times \notag \\
    &\quad \; \frac{(2c_{k,2(t-1)+1}-1)(\varsigma_t^2\mbf{\hat{g}}_{k,2(t-1)+1} -  \varepsilon^2\hat{z}_{k,t}^\Re \mbf{U}_t \mbf{\tilde{x}}) )}{\sqrt{\pi}\varsigma_t^{3}},\notag \\
    &\nabla_{\mbf{\tilde{x}}}\, p_{k,t}^{\Re,e_1} = -\nabla_{\mbf{\tilde{x}}}\, p_{k,t}^{\Re,e_0},\notag \\
    &\nabla_{\mbf{\tilde{x}}}\, p_{k,t}^{\Im,e_0} = -e^{-(\hat{z}_{k,t}^\Im)^2/\varsigma_t^2} \; \times \notag \\
    &\quad \; \frac{(2c_{k,2(t-1)+2}-1)(\varsigma_t^2\mbf{\hat{g}}_{k,2(t-1)+2} -  \varepsilon^2\hat{z}_{k,t}^\Im \mbf{U}_t \mbf{\tilde{x}}) )}{\sqrt{\pi}\varsigma_t^{3}},\notag \\
    &\nabla_{\mbf{\tilde{x}}}\, p_{k,t}^{\Im,e_1} = -\nabla_{\mbf{\tilde{x}}}\, p_{k,t}^{\Im,e_0},\notag 
\end{align}
where $\mbf{U}_t$ is a diagonal matrix whose diagonal elements are zero except those from position $(2(t-1)N+1)$ to $(2tN)$, and
\begin{align*}
    [\mbf{\hat{g}}_{k,1},\,\ldots,\,\mbf{\hat{g}}_{k,2\Ls}]^T = \mbf{I}_{\Ls}\otimes \begin{bmatrix}
        \Re\{\mbf{\hat{h}}_k^T\} & -\Im\{\mbf{\hat{h}}_k^T\}
        \\
        \Im\{\mbf{\hat{h}}_k^T\} & \Re\{\mbf{\hat{h}}_k^T\}
    \end{bmatrix}.
\end{align*}
A similar derivation can be performed for 8-QAM and 16-QAM, although for brevity we omit the details.

\begin{figure}[t!]
    \centering
    \includegraphics[width=0.85\linewidth]{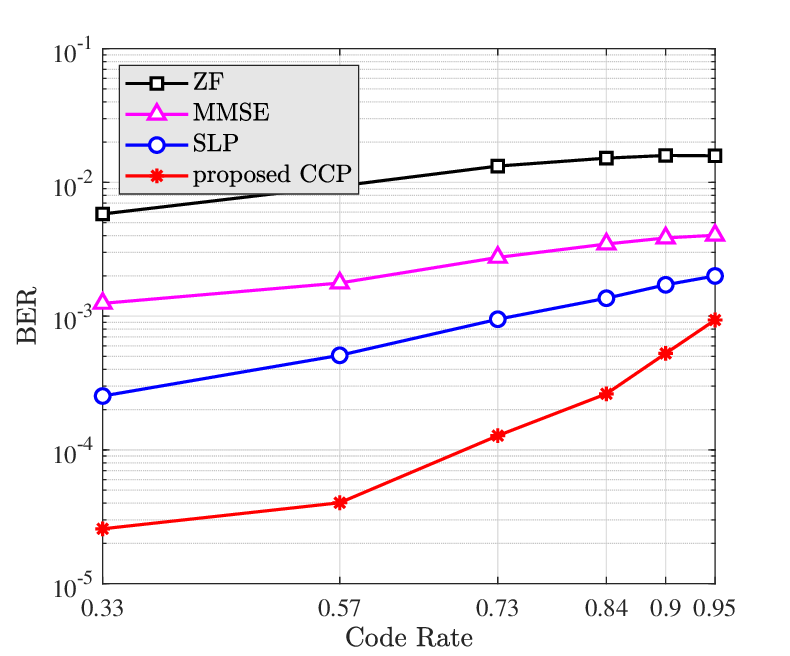}
    \caption{BER versus code-rate performance for $K = N = 4$, $4$-QAM, $\ec = 1$, and $\mrm{SNR} = 15$dB. Hamming codes [3,1], [7,4], [15,11], [31,26], [63,57], [127,120] are used to generate different code rates.}
    \label{fig:BER_vs_codeRate}
\end{figure}
\begin{figure*}[t!]
    \centering
    \begin{subfigure}[t]{0.24\linewidth}
        \centering
        \includegraphics[width=\linewidth]{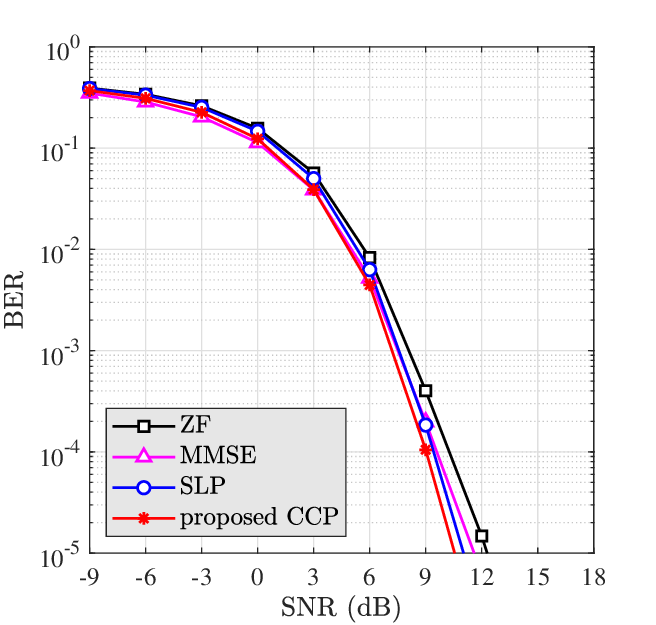}
        \caption{$K=3$ users.}
        \label{fig:BER_vs_SNR_Hamming74_3K_6N}
    \end{subfigure}~
    \begin{subfigure}[t]{0.24\linewidth}
        \centering
        \includegraphics[width=\linewidth]{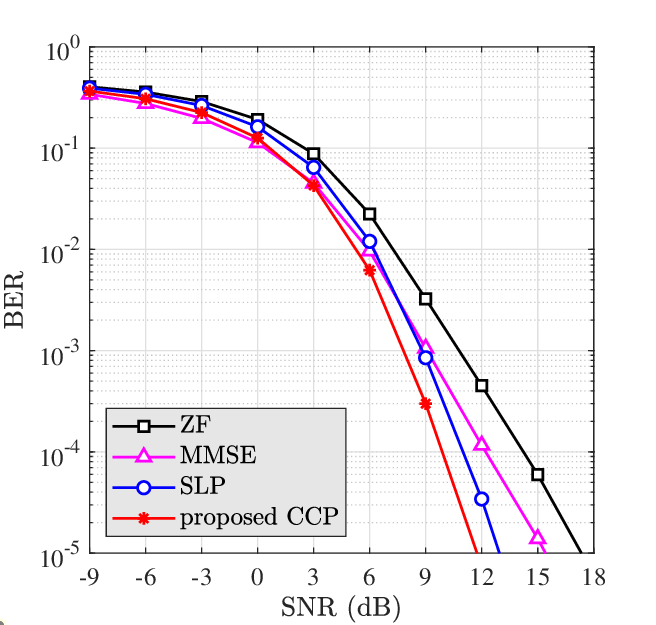}
        \caption{$K=4$ users.}
        \label{fig:BER_vs_SNR_Hamming74_4K_6N}
    \end{subfigure}~
    \begin{subfigure}[t]{0.24\linewidth}
        \centering
        \includegraphics[width=\linewidth]{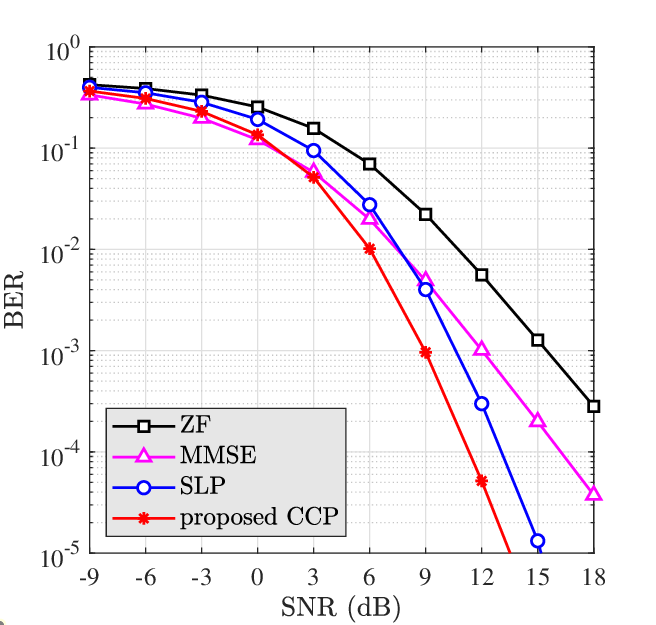}
        \caption{$K=5$ users.}
        \label{fig:BER_vs_SNR_Hamming74_5K_6N}
    \end{subfigure}~
    \begin{subfigure}[t]{0.24\linewidth}
        \centering
        \includegraphics[width=\linewidth]{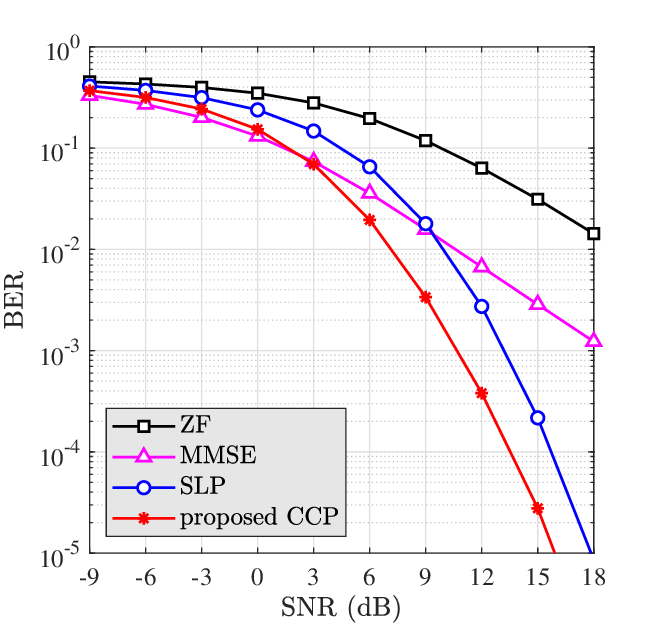}
        \caption{$K=6$ users.}
        \label{fig:BER_vs_SNR_Hamming74_6K_6N}
    \end{subfigure}
    \caption{BER performance for different number of users $K$ with $N=6$ transmit antennas, 4-QAM, and $\ec = 1$ (Hamming [7,4]).}
    \label{fig:performance_Hamming74_6N}
\end{figure*}
\begin{figure*}[t!]
    \centering
    \begin{subfigure}[t]{0.32\linewidth}
        \centering
        \includegraphics[width=\linewidth]{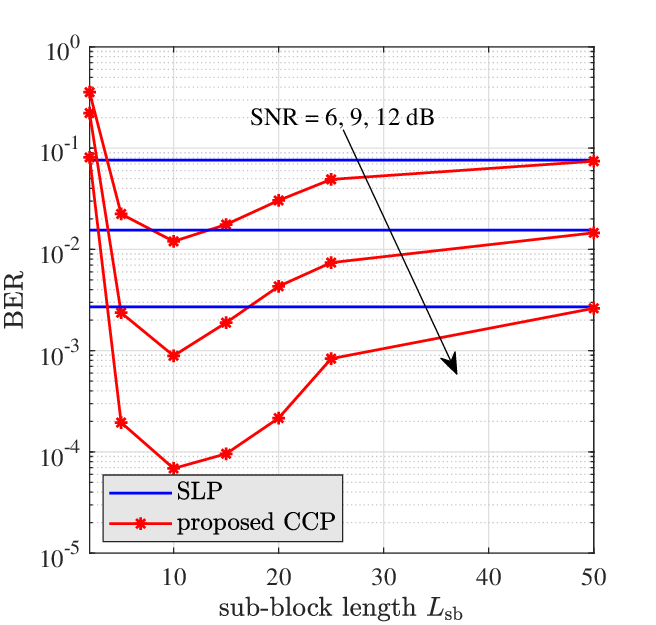}
        \caption{4-QAM.}
        \label{fig:BER_vs_Lsb_convolutional_4QAM}
    \end{subfigure}~
    \begin{subfigure}[t]{0.32\linewidth}
        \centering
        \includegraphics[width=\linewidth]{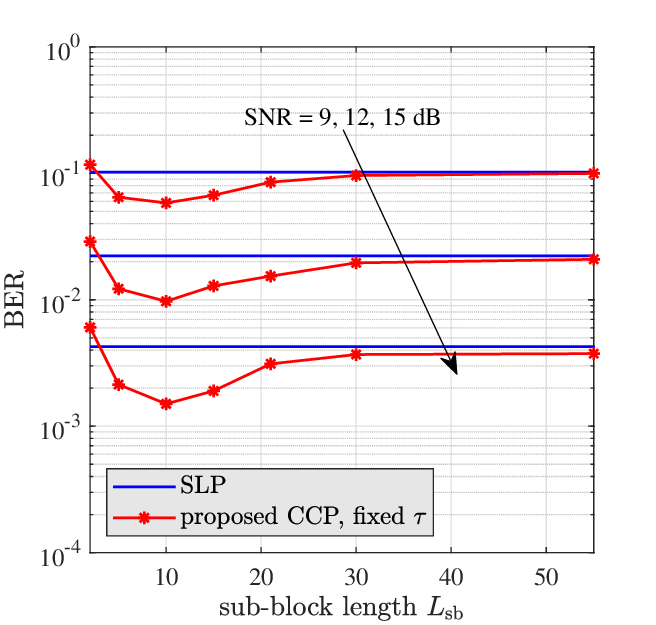}
        \caption{8-QAM.}
        \label{fig:BER_vs_Lsb_convolutional_8QAM}
    \end{subfigure}~
    \begin{subfigure}[t]{0.32\linewidth}
        \centering
        \includegraphics[width=\linewidth]{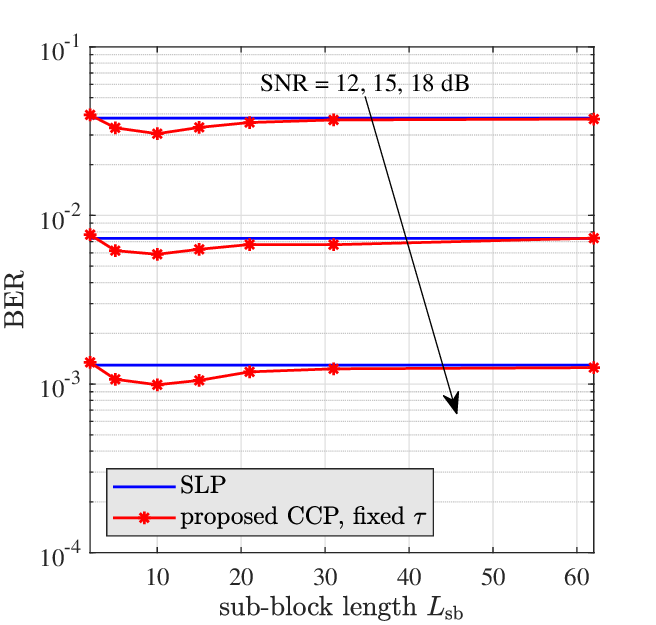}
        \caption{16-QAM.}
        \label{fig:BER_vs_Lsb_convolutional_16QAM}
    \end{subfigure}
    \caption{BER versus sub-block-length for different constellations with $K=N=4$ and convolutional code ($\ec > 1$).}
    \label{fig:BER_vs_Lsb_convolutional}
\end{figure*}

\section{Numerical Results}
\label{sec:numerical_results}
In this section, we provide simulation results to {assess the performance of the proposed CCP method and compare it against the different conventional precoding approaches mentioned in Section~\ref{subsec:conventional_precoders}}. We model the channel to the $k$-th user as $\mbf{h}_k \sim \mca{CN}(0, \gamma_k\mbf{I}_N)$ where $\gamma_k = \gamma_0 d_k^{-\nu}$ represents the large scale fading coefficient. We set $\gamma_0 = -30$ dB as the reference path loss and $\nu = 2.6$ as the path loss exponent. The distance $d_k$ between user $k$ and the BS is randomly generated between 200 and 300m. For implementing Algorithm~\ref{algo:gradCCP}, we set $\alpha_0 = 1$, $\alpha_{\mrm{min}} = 10^{-5}$, $\eta = 0.95$, $\kappa_{\mrm{max}} = 50$, and $\ell_{\mrm{max}} = 5000$. For implementing Algorithm~\ref{algo:gradCCP_tau_design}, we set $\zeta = 1.05$ and $\xi = 0.99$. We use conventional SLP to initialize the algorithms. When CCP is implemented with a fixed decision threshold $\tau$, the threshold given by SLP is used. In our simulations, we use either Hamming codes or convolutional codes with code rate $1/2$ for the case of $\ec = 1$ and $\ec > 1$, respectively. The employed convolutional code has a free distance $d_{\mrm{free}} = 10$, which gives $\ec = \lfloor(d_{\mrm{free}}-1)/2\rfloor = 4$. We set the length of the information bit sequence as $\Lb = 498$ bits. If not explicitly stated, hard channel decoding and perfect CSI are assumed. We use the same SNR definition as in~\cite{Junwen2023Speeding} which is given by $\mrm{SNR} = \|\mbf{HX}\|_F^2/(K\Ls\sigma^2)$. Note that all of the BER results shown in this section {refer to the error rate of the information bits.}

Fig.~\ref{fig:performance_Hamming74_4K_4N} provides BER performance and noiseless received signal comparisons between the proposed CCP and several other existing approaches for the case of $\ec = 1$. The sample plot of noiseless received signals in Fig.~\ref{fig:noiseless_rx_signal} illustrates the key idea behind CCP. Unlike the conventional SLP approach that forces all received signals to strictly reside within the desired constructive interference regions, the proposed CCP design allows for some erroneous received symbols since they can be corrected by the channel code. Aside from these symbol errors, the remaining symbols designed by CCP are positioned farther from the decision boundaries compared to symbols designed by SLP, making them more resilient to noise. Since the bit errors resulting from the erroneous received symbols can be corrected by the channel decoder, CCP will have superior information BER performance, as seen from the results in Fig.~\ref{fig:BER_vs_SNR_Hamming74_4K_4N}. In this scenario, CCP provides a gain of over 3 dB compared to the SLP approach. {This demonstrates that the proposed optimization metric in~\eqref{eq:problem1} is appropriate, taking into account the error-correcting capability of the channel code to design efficient transmit signals.}

Next, we evaluate the effect of the code rate in Fig.~\ref{fig:BER_vs_codeRate}, which shows that the performance gain of CCP is more significant for lower code rates. This is expected because Hamming codes with lower code rates have shorter codewords, allowing for more symbol errors and providing greater flexibility in signal design. More generally, channel codes with a higher error-correcting capacity allow CCP to introduce more symbol errors while at the same time leveraging the increased flexibility gained from these errors to boost the power of the received signals for the other symbols. In Fig.~\ref{fig:performance_Hamming74_6N}, we compare the performance of CCP with other existing approaches as the number of users $K$ increases for a fixed number of transmit antennas. It is observed that the performance gain of CCP increases as the system becomes more loaded. This is due to the fact that as $K$ increases, more multiuser interference is present, which together with the ability to correct symbol errors provides more degrees of freedom to push the received signal away from the decision boundaries. A similar observation has also been made in the SLP literature when comparing SLP with other conventional precoding approaches. 

\begin{figure*}[t!]
    \centering
    \begin{subfigure}[t]{0.32\linewidth}
        \centering
        \includegraphics[width=\linewidth]{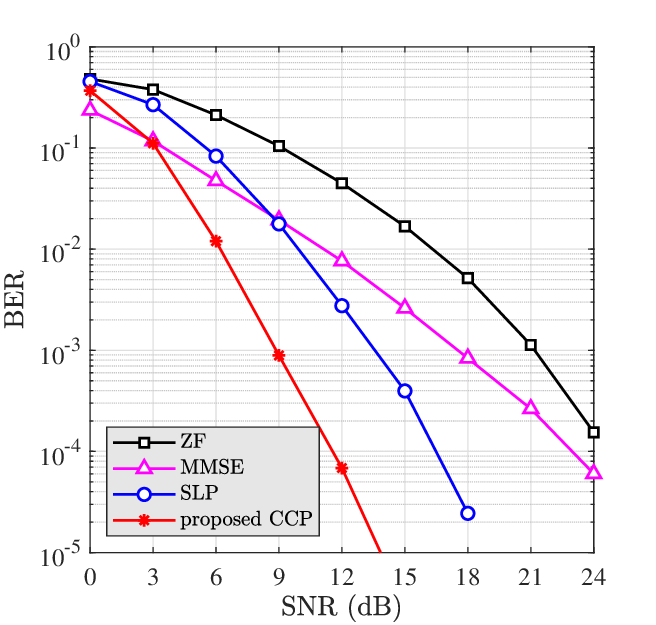}
        \caption{4-QAM.}
        \label{fig:BER_vs_SNR_convolutional_4QAM}
    \end{subfigure}~
    \begin{subfigure}[t]{0.32\linewidth}
        \centering
        \includegraphics[width=\linewidth]{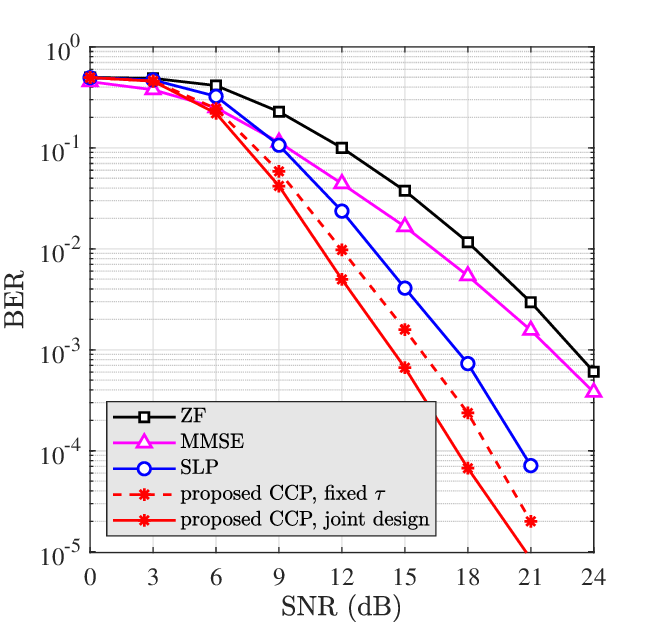}
        \caption{8-QAM.}
        \label{fig:BER_vs_SNR_convolutional_8QAM}
    \end{subfigure}~
    \begin{subfigure}[t]{0.32\linewidth}
        \centering
        \includegraphics[width=\linewidth]{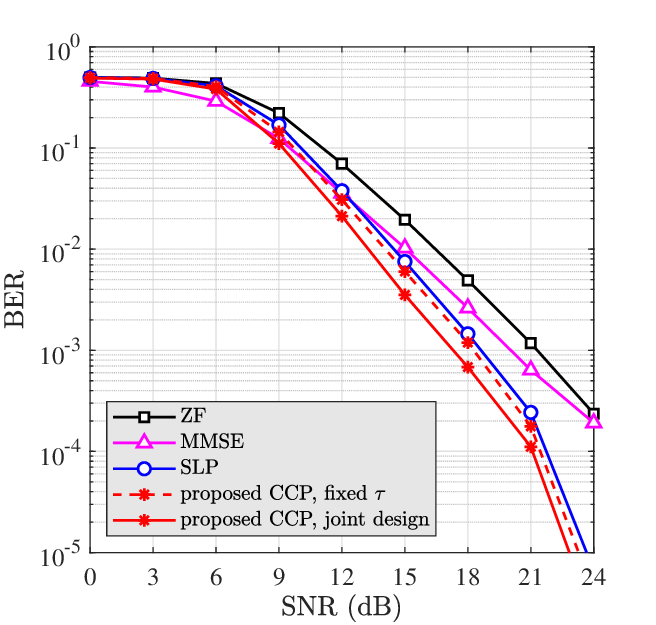}
        \caption{16-QAM.}
        \label{fig:BER_vs_SNR_convolutional_16QAM}
    \end{subfigure}
    \caption{BER versus SNR performance for different constellations with $K=N=4$, convolutional code ($\ec > 1$), and sub-block length $L_{\mrm{sb}} = 10$.}
    \label{fig:BER_vs_SNR_convolutional}
\end{figure*}

In Figs.~\ref{fig:BER_vs_Lsb_convolutional} and~\ref{fig:BER_vs_SNR_convolutional}, we consider a system with $\ec > 1$ that employs a convolutional code. Since the proposed CCP approach for $\ec > 1$ is based on dividing the transmit signal into sub-blocks, we investigate the effect of the sub-block length, $\Lsb$, in Fig.~\ref{fig:BER_vs_Lsb_convolutional}. It can be seen that there is an optimal sub-block length for which the CCP achieves its best performance. Both excessively short and long sub-block lengths degrade the CCP performance. This is because smaller values of $\Lsb$ increase the number of sub-blocks, which inevitably leads to CCP allowing and thus increasing the likelihood that the number of bit errors at the receiver will exceed the channel code’s error-correcting capacity. Conversely, larger values of $\Lsb$ reduce the number of sub-blocks, and hence limits the ability of CCP to fully exploit the error-correcting capacity of the channel code, as fewer symbol errors can be introduced by CCP. Therefore, selecting an appropriate sub-block length is critical for obtaining the best performance. For example, in this scenario, a sub-block length of $\Lsb = 10$ is ideal.

BER-versus-SNR performance comparisons for 4-QAM, 8-QAM, and 16-QAM constellations with a convolutional code are given in Fig.~\ref{fig:BER_vs_SNR_convolutional} when $\Lsb=10$. The results demonstrate that CCP outperforms the existing methods, particularly for smaller constellations like 4-QAM, where a gain of up to 5 dB is observed. However, as the constellation size increases, the performance gain for CCP becomes smaller. This is because larger constellations reduce the size of the constructive interference regions, leading to a lower chance of benefiting from multiuser interference introduced by the symbol errors in the CCP design. The results in Figs.~\ref{fig:BER_vs_SNR_convolutional_8QAM} and~\ref{fig:BER_vs_SNR_convolutional_16QAM} also demonstrate the benefit of the CCP approach proposed in Algorithm~\ref{algo:gradCCP_tau_design} where the transmit signal $\mbf{X}$ and decision threshold $\tau$ are jointly optimized. 

\begin{figure}[t!]
    \centering
    \includegraphics[width=0.8\linewidth]{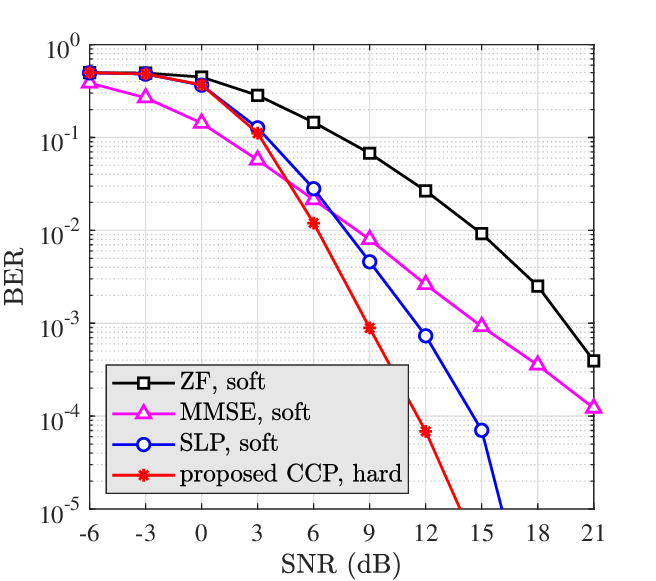}
    \caption{BER performance comparison between CCP with hard decoding and other methods with soft decoding, $K = N = 4$, $4$-QAM, and convolutional code with sub-block length $L_{\mrm{sb}} = 10$.}
    \label{fig:soft_BER}
\end{figure}

We compare the performance of CCP with other conventional approaches with soft decoding in Fig.~\ref{fig:soft_BER}, illustrating in this example that CCP with hard decoding outperforms conventional approaches that utilize soft decoding. This result suggests that with well-designed transmit signals, receivers utilizing low-complexity hard decoding can surpass the performance of conventional methods with computationally intensive soft decoding techniques. Therefore, the proposed CCP approach is well-suited for systems with receivers constrained by limited power and computational resources.

\begin{figure}[t!]
    \centering
    \includegraphics[width=0.8\linewidth]{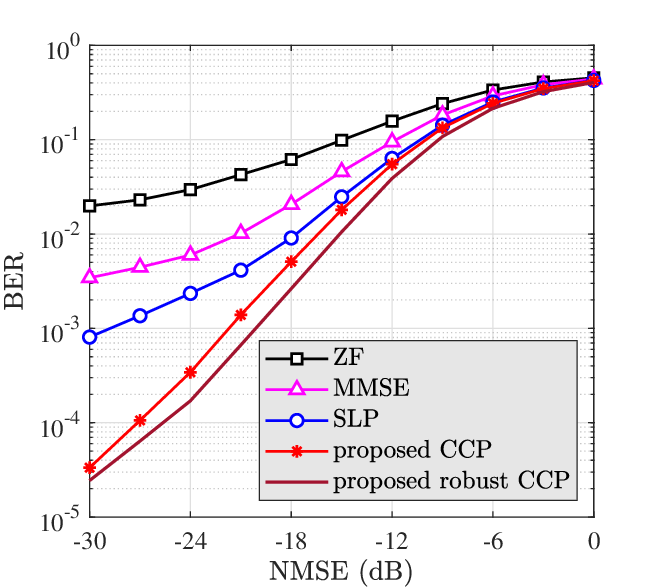}
    \caption{BER performance comparison for imperfect CSI, $K = N = 4$, $4$-QAM, and convolutional code with sub-block length $L_{\mrm{sb}} = 10$.}
    \label{fig:robust_BER}
\end{figure}

In Fig.~\ref{fig:robust_BER}, we evaluate the performance of the version of CCP designed for robustness to channel estimation error. We assume the imperfect CSI model discussed in Section~\ref{sec:robust_CCP} and we set the variance of the channel estimation error $\varepsilon^2$ to match the following definition of normalized mean-squared error (NMSE): $\text{NMSE} = KN\varepsilon^2/\|\mbf{H}\|_F^2$. It can be seen that the BER of the proposed robust CCP approach is about 2dB lower than the version of CCP that ignores channel estimation errors, illustrating the effectiveness of taking the CSI errors into account.

\section{Conclusion}
\label{sec:conclusion}
In this paper, we have proposed a novel precoding framework for multi-user coded systems, referred to as Channel-Coded Precoding. The proposed CCP framework {takes into account the channel coding} and aims to minimize the ultimate BER of interest in coded systems, which is the BER of the information bits. Unlike prior work, our approach maximizes the probability that the {\em information bits} rather than that coded bits are correctly recovered by the channel decoder, a modification that {provides} more degrees of freedom for {enhanced} transmit signal designs, thereby significantly improving the overall system performance. CCP designs for both cases involving one-bit and multi-bit error-correcting capacity have been proposed. We have also developed a robust CCP approach for the case where perfect CSI is unavailable. Our results suggest that the performance gains of CCP are most significant in systems with low-order constellations, lower code rates, and where the number of users is comparable to the number of transmit antennas. The novel precoding perspective described in the paper introduces a new approach for transmit signal design in which channel coding and multi-antenna precoding are jointly considered/optimized.


\appendices
\section{Calculation of the Gradients 
for 8-QAM and 16-QAM.}
\label{appendix_1}
For the 8-QAM constellation, the gradients $\nabla_{\mbf{\bar{x}}}\, p_{k,t}^{\Re,e_0}$ and $\nabla_{\mbf{\bar{x}}}\, p_{k,t}^{\Re,e_1}$ are given as follows:

\textbf{Case 1:} $\{c_{k,3(t-1)+1}, c_{k,3(t-1)+2}\} = \{0,0\}$.
\begin{align}
    \nabla_{\mbf{\bar{x}}}\, p_{k,t}^{\Re,e_0} &= \frac{-1}{\sigma\sqrt{\pi}}\begin{bmatrix}[0.6]
            1\\\mbf{g}_{k,2(t-1)+1}
        \end{bmatrix}\,e^{-(\tau + z_{k,t}^\Re)^2/\sigma^2}. \\
    \nabla_{\mbf{\bar{x}}}\, p_{k,t}^{\Re,e_1} &= \frac{-1}{\sigma\sqrt{\pi}}\begin{bmatrix}[0.6]
            0\\\mbf{g}_{k,2(t-1)+1}
        \end{bmatrix}\,e^{-(z_{k,t}^\Re)^2/\sigma^2} \; + \notag \\
        &\quad \; \frac{1}{\sigma\sqrt{\pi}}\begin{bmatrix}[0.6]
            1\\\mbf{g}_{k,2(t-1)+1}
        \end{bmatrix}\,e^{-(\tau + z_{k,t}^\Re)^2/\sigma^2} \; + \notag \\
        &\quad \; \frac{1}{\sigma\sqrt{\pi}}\begin{bmatrix}[0.6]
            -1\\\mbf{g}_{k,2(t-1)+1}
        \end{bmatrix}\,e^{-(-\tau + z_{k,t}^\Re)^2/\sigma^2}.
\end{align}

\textbf{Case 2:} $\{c_{k,3(t-1)+1}, c_{k,3(t-1)+2}\} = \{0,1\}$.
\begin{align}
    \nabla_{\mbf{\bar{x}}}\, p_{k,t}^{\Re,e_0} &= \frac{-1}{\sigma\sqrt{\pi}}\begin{bmatrix}[0.6]
            0\\\mbf{g}_{k,2(t-1)+1}
        \end{bmatrix}\,e^{-(z_{k,t}^\Re)^2/\sigma^2} \; + \notag \\
        &\quad \; \frac{1}{\sigma\sqrt{\pi}}\begin{bmatrix}[0.6]
            1\\\mbf{g}_{k,2(t-1)+1}
        \end{bmatrix}\,e^{-(\tau + z_{k,t}^\Re)^2/\sigma^2}. \\
    \nabla_{\mbf{\bar{x}}}\, p_{k,t}^{\Re,e_1} &=  \frac{-1}{\sigma\sqrt{\pi}}\begin{bmatrix}[0.6]
            1\\\mbf{g}_{k,2(t-1)+1}
        \end{bmatrix}\,e^{-(\tau + z_{k,t}^\Re)^2/\sigma^2} \; + \notag \\
        &\quad \; \frac{-1}{\sigma\sqrt{\pi}}\begin{bmatrix}[0.6]
            -1\\\mbf{g}_{k,2(t-1)+1}
        \end{bmatrix}\,e^{-(-\tau + z_{k,t}^\Re)^2/\sigma^2} \; + \notag \\
        &\quad \; \frac{1}{\sigma\sqrt{\pi}}\begin{bmatrix}[0.6]
            0\\\mbf{g}_{k,2(t-1)+1}
        \end{bmatrix}\,e^{-(z_{k,t}^\Re)^2/\sigma^2}.
\end{align}

\textbf{Case 3:} $\{c_{k,3(t-1)+1}, c_{k,3(t-1)+2}\} = \{1,0\}$.
\begin{align}
    \nabla_{\mbf{\bar{x}}}\, p_{k,t}^{\Re,e_0} &=  \frac{1}{\sigma\sqrt{\pi}}\begin{bmatrix}[0.6]
            -1\\\mbf{g}_{k,2(t-1)+1}
        \end{bmatrix}\,e^{-(-\tau + z_{k,t}^\Re)^2/\sigma^2}. \\
    \nabla_{\mbf{\bar{x}}}\, p_{k,t}^{\Re,e_1} &=  \frac{-1}{\sigma\sqrt{\pi}}\begin{bmatrix}[0.6]
            1\\\mbf{g}_{k,2(t-1)+1}
        \end{bmatrix}\,e^{-(\tau + z_{k,t}^\Re)^2/\sigma^2} \; + \notag \\
        &\quad \; \frac{-1}{\sigma\sqrt{\pi}}\begin{bmatrix}[0.6]
            -1\\\mbf{g}_{k,2(t-1)+1}
        \end{bmatrix}\,e^{-(-\tau + z_{k,t}^\Re)^2/\sigma^2} \; + \notag \\
        &\quad \; \frac{1}{\sigma\sqrt{\pi}}\begin{bmatrix}[0.6]
            0\\\mbf{g}_{k,2(t-1)+1}
        \end{bmatrix}\,e^{-(z_{k,t}^\Re)^2/\sigma^2}.
\end{align}

\textbf{Case 4:} $\{c_{k,3(t-1)+1}, c_{k,3(t-1)+2}\} = \{1,1\}$.
\begin{align}
    \nabla_{\mbf{\bar{x}}}\, p_{k,t}^{\Re,e_0} &= \frac{-1}{\sigma\sqrt{\pi}}\begin{bmatrix}[0.6]
            -1\\\mbf{g}_{k,2(t-1)+1}
        \end{bmatrix}\,e^{-(-\tau + z_{k,t}^\Re)^2/\sigma^2} \; + \notag\\
        &\quad \; \frac{1}{\sigma\sqrt{\pi}}\begin{bmatrix}[0.6]
            0\\\mbf{g}_{k,2(t-1)+1}
        \end{bmatrix}\,e^{-(z_{k,t}^\Re)^2/\sigma^2.} \\
    \nabla_{\mbf{\bar{x}}}\, p_{k,t}^{\Re,e_1} &=  \frac{-1}{\sigma\sqrt{\pi}}\begin{bmatrix}[0.6]
            0\\\mbf{g}_{k,2(t-1)+1}
        \end{bmatrix}\,e^{-(z_{k,t}^\Re)^2/\sigma^2} \; + \notag \\
        &\quad \; \frac{1}{\sigma\sqrt{\pi}}\begin{bmatrix}[0.6]
            1\\\mbf{g}_{k,2(t-1)+1}
        \end{bmatrix}\,e^{-(\tau + z_{k,t}^\Re)^2/\sigma^2} \; + \notag \\
        &\quad \; \frac{1}{\sigma\sqrt{\pi}}\begin{bmatrix}[0.6]
            -1\\\mbf{g}_{k,2(t-1)+1}
        \end{bmatrix}\,e^{-(-\tau + z_{k,t}^\Re)^2/\sigma^2}.
\end{align}
The gradients $\nabla_{\mbf{\bar{x}}}\, p_{k,t}^{\Im,e_0}$ and $\nabla_{\mbf{\bar{x}}}\, p_{k,t}^{\Im,e_1}$ for 8-QAM are similar to those in~\eqref{eq:grad_pIm0_4QAM} and~\eqref{eq:grad_pIm1_4QAM}, respectively, since the imaginary part of 8-QAM symbols only carries one bit of information like 4-QAM. For the case of 16-QAM, the gradients can be obtained in the same way as the four cases above since both the real and imaginary parts of the 16-QAM symbols carry two bits of information.

\bibliographystyle{IEEEtran}
\bibliography{ref}

%









\end{document}